\documentclass[showpacs,twocolumn,aps,preprintnumbers,letterpaper]{revtex4}
\usepackage{amsmath,amssymb}
\usepackage{epsfig}
\usepackage{graphicx}
\usepackage{amsmath}
\usepackage{slashed}
\usepackage{amsfonts}
\usepackage{epstopdf}
\usepackage{color}

\newcommand{\be}{\begin{equation}}
\newcommand{\ee}{\end{equation}}
\newcommand{\bear}{\begin{eqnarray}}
\newcommand{\eear}{\end{eqnarray}}
\newcommand{\ba}{\begin{array}}
\newcommand{\ea}{\end{array}}

\def\be{\begin{eqnarray}}
\def\ee{\end{eqnarray}}
\def\bea{\be}
\def\eea{\ee}

\def\roughly#1{\mathrel{\raise.3ex\hbox{$#1$\kern-.75em%
\lower1ex\hbox{$\sim$}}}}

\begin{document}

\title{Rotating Dirac fermions in a  magnetic field in 1+2,3 dimensions}

\author{Yizhuang Liu  and Ismail Zahed}
\email{yizhuang.liu@stonybrook.edu}
\email{ismail.zahed@stonybrook.edu}
\affiliation{Department of Physics and Astronomy, Stony Brook University, Stony Brook, New York 11794-3800, USA}



\date{\today}
\begin{abstract}
We consider the effects of an external magnetic field on rotating fermions in 1+2,3 dimensions.
The dual effect of a rotation parallel to the magnetic field causes a net increase in the fermionic density
by centrifugation, which follows from the sinking of the particle lowest Landau level in the Dirac sea for
free Dirac fermions. In 1+d = 2n dimensions,  this effect  is related to the chiral magnetic
effect in 2n-2 dimensions. This phenomenon is  discussed specifically for both weak and strong inter-fermion
interactions in 1+2 dimensions.  For QCD in 1+3 dimensions with Dirac quarks, we show that in the strongly coupled phase
with spontaneously broken chiral symmetry, this mechanism reveals itself in the form of an induced pion
condensation by centrifugation. We use this observation to show that this effect causes a shift in the
chiral condensate in leading order in the pion interaction, and to discuss the possibility for the formation of a novel pion super-fluid phase in off-central heavy ion collisions at collider energies.
\end{abstract}


\maketitle

\setcounter{footnote}{0}


\section{Introduction}

The combined effects of rotations and magnetic fields on  Dirac fermions are realized in a
wide range of physical settings ranging from macroscopic spinning neutron stars and black holes
\cite{VILENKIN}, all the way to
microscopic anomalous transport in Weyl metals~\cite{WEYL}. In any dimensions, strong magnetic fields
reorganize the fermionic spectra into Landau levels, each with a huge planar degeneracy that
is lifted when a paralell rotation is applied. The past decade has seen a large interest in the
chiral and vortical effects and their relationship with anomalies~\cite{ANOMALY} (and references therein).

Perhaps, a less well known effect stems from the dual combination of a rotation and magnetic
field on free or interacting Dirac fermions. Recently, it was noted  that this dual combination
could lead to novel effects for composite fermions at half filling in 1+2 dimensions under the
assumption that {they are Dirac fermions}~\cite{US},  and more explicitly for free and interacting
Dirac fermions in 1+3  dimensions~\cite{YIN,LIAO,FUKU}.  Indeed, when a rotation is applied along a magnetic field,
the charge density was observed to increase {\it in the absence} of a chemical potential.
A possible relationship of this phenomenon to
the Chern-Simons term in odd dimensions, and the chiral anomaly in even dimensions was
suggested.

The purpose of this paper is to revisit these issues in a more explicit way in 1+2,3 dimensions.
The case of 1+2 dimensions is of interest to planar materials in the context of solid state
physics, while the case of 1+3 dimensions is of more general interest with relation to QCD.
Recently, there have been few studies along these lines using effective models of the NJL type
in 1+3 dimensions, where the phenomenon of charge density enhancement was also confirmed
with new observations~\cite{LIAO,FUKU}. Also, recent analyses using pion effective descriptions 
have suggested the possibility of Bose condensation in strong magnetic fields~\cite{AYA} and 
dense matter with magnetism or rotations~\cite{YAMA}.

 This paper consists of a number of new results:
 1/ a full analysis of the combined effects of a rotation and magnetic field
 on free and interacting Dirac fermions in 1+2 dimensions, both at weak and strong coupling;
 2/ a correspondence with anomalies in arbitrary dimensions;
 3/ a deformation of the current densities by centrifugation  in the presence of a magnetic field;
 4/  a depletion of the QCD chiral condensate in leading order in the pion interaction;
 5/ a charge pion condensation induced by centrifugation in a magnetic field.

The outline of the paper is as follows:
In section II we detail the Landau level problem for free Dirac fermions in 1+2 dimensions in the presence
of an arbitrary rotation described using  a local metric. In section III we explore the effects of  the interaction on
the free results through a 4-Fermi interaction
 both in the weak and strong coupling regime. In section IV and V we extend our chief observations
to 1+3 dimensions to the free and interacting fermionic cases
with particular interest to the shift in the chiral condensate in QCD. In section VI, we discuss the possibility
for the formation of a pion BEC phase in off-central heavy ion collisions.
Our conclusions are in section VII. We record in the Appendices useful details regarding some of the calculations.

\section{Dirac fermions in  1+2}

In this section we will outline how to implement a global rotation through a pertinent metric. We will then
use it to derive explicit results for massless Dirac fermions with a global $U(2)$ symmetry in the presence
of a parallel magnetic field in 1+2 dimensions. The basic mechanism of the shift caused by the rotation on the
LLL will be clearly elucidated, and both the scalar and vector densities evaluated.

\subsection{Metric for a rotating frame}

To address the effects of a finite rotation $\Omega$  in $1+2$ dimensions we define  the rotating metric

\be
\label{1}
ds^2=(1-\Omega^2\rho^2)dt^2+2y\Omega dxdt-2x\Omega dydt
\ee
The frame fields or veilbeins are defined as $g^{\mu\nu}=e^\mu_ae^\nu_a\eta_{ab}$ with signature
$\sqrt{-g}=1$, in terms of which the co-moving frame is $\theta^{a}=e^{a}_{\mu}dx^{\mu}$ and
$e_{a}=e^{\mu}_a\partial_{\mu}$ are explicitly given by

\be
\label{2}
(\theta^0,\theta^1,\theta^2)=(dt, dx-y\Omega dt, dy+x\Omega dt)\nonumber\\
(e_0, e_1, e_2)=(\partial_t+y\Omega \partial_x-x\Omega \partial_y, \partial_1, \partial_2)
\ee
with the spin connections

\bea
\label{3}
&&\omega^{1}_0=\omega_1^{0}=+\Omega(dy-\Omega xdt)\nonumber\\
&&\omega^{2}_0=\omega_2^{0}=-\Omega(dx+\Omega ydt)
\eea
In a fixed area of size $S=\pi R^2$, the time-like nature of the metric (\ref{1})
and therefore causality are maintained for  $\Omega R\leq 1$. The importance
of a finite size for rotating fermions was emphasized in~\cite{FUKU}. This will be
understood throughout.

\subsection{Rotation plus magnetic field}

The Lagrangian that describes free rotating Dirac fermions in a fixed magnetic field in $1+2$ dimensions, reads

\bea
\label{4}
{\cal L}&&=\bar \psi(i\gamma^{\mu}(D_\mu+\Gamma_\mu)-M)\psi\nonumber \\
&&=\bar \psi(i\gamma^0(\partial_t-\Omega(x\partial_y-y\partial_x+iS^z))+i\gamma^{i}D_i-M)\psi\nonumber\\
\eea
with the long derivative $D=\partial-ieA$, and the following choice of gamma matrices,
$\gamma^{a}$ as $\gamma^0={\rm diag}(\sigma_3,-\sigma_3)$,$\gamma^1={\rm diag}(i\sigma_1,-i\sigma_1)$,$\gamma^2
={\rm diag}(i\sigma_2,-i\sigma_2)$, to accomodate for both particles and anti-particles.

A thorough  analysis of (\ref{4})  for
an external vector potential in a rotationally non-symmetric gauge was given in~\cite{MIRANSKY}. Here we insist on preserving
rotational symmetry by choosing $A_{\mu}=(0,By/2,-Bx/2,0)$. As a result, the LL spectrum is characterized explicitly by both
energy and angular momentum conservation which are described in terms of the anti-commutative harmonic oscillator $a, b$ operators

\bea
\label{5}
a&&=\frac{i}{\sqrt{2eB}}(D_x+iD_y)=-\frac{i}{\sqrt{2eB}}\left(2\bar \partial+\frac{eBz}{2} \right)\nonumber\\
b&&=\frac{1}{\sqrt{2eB}}\left(2\partial+\frac{eB\bar z}{2} \right)
\eea
Throughout, we will assume $eB>0$ unless specified otherwise.
The rotating Landau levels are  labelled by $m,n$ as

\be
\label{6}
E^{\pm}+\Omega(m-n+\frac{1}{2})=\pm\sqrt{M^2+2eBn}=\pm \tilde E
\ee
for particles and anti-particles.
The corresponding normalized scalar wave functions for the n-th
Landau level  with good angular momentum $l_z=xp_y-yp_z=b^{\dagger}b-a^{\dagger}a$ with
eigenvalue $m-n$, are

\be
\label{7}
f_{nm}=\frac{(a^{\dagger})^n(b^{\dagger})^m}{\sqrt{n!m!}}f_{00}
\ee
with the lowest Landau level (LLL) $f_{00}\propto e^{-\frac 14 eB (x^2+y^2)}$.
Note  that for n=0, we have only one positive energy state with spin up, and one negative energy state
with spin down, each with degeneracy $N=eBS/2\pi$.
For $\Omega=0$ and $n>0$ all Landau level (LL)  have degeneracy $2N=eBS/\pi$. The degeneracy
is lifted by centrifugation for $\Omega\neq 0$.

In terms of (\ref{7}) the quantized Dirac fields
follow in the form

\be
\label{8}
\psi(t,\vec x)=\sum_{nmi}( u_{nm}^i(\vec x)e^{-iE^{+}t}\,a_{nm}^i +v_{nm}^i (\vec x)e^{-iE^{-}t}\,b^{i\dagger }_{nm})\nonumber\\
\ee
where $a^i_{nm}$ annihilates a particle with positive energy $E^+$ and spin $i=\pm \frac 12$, and
$b^{i\dagger}_{nm}$ creates a hole with negative energy $E^-$ and spin $i=\mp \frac 12$.
Their corresponding wavefunctions are

\bea
\label{9}
u_{0m}&&=(f_{0m},0,0,0)\nonumber\\
v_{0m}&&=(0,0,f_{0m},0)\nonumber\\
u_{nm}^{+}&&=\sqrt{\frac{\tilde  E+M}{2\tilde E}}\left(f_{nm},\frac{i\sqrt{2eB}}{\tilde E+M}f_{n-1,m},0,0\right)\nonumber\\
u_{nm}^{-}&&=\sqrt{\frac{\tilde  E-M}{2\tilde E}}\left(0,0,f_{nm},-\frac{i\sqrt{2eB}}{\tilde E-M}f_{n-1,m}\right)\nonumber\\
v_{nm}^{+}&&=\sqrt{\frac{\tilde  E-M}{2\tilde E}}\left(f_{nm},-\frac{i\sqrt{2eB}}{\tilde E-M}f_{n-1,m},0,0\right)\nonumber\\
v_{nm}^{-}&&=\sqrt{\frac{\tilde  E+M}{2\tilde E}}\left(0,0,f_{nm},\frac{i\sqrt{2eB}}{\tilde E+M}f_{n-1,m}\right)
\eea

\subsection{Scalar density}

\begin{figure}[t]
  \begin{center}
  \includegraphics[width=8cm]{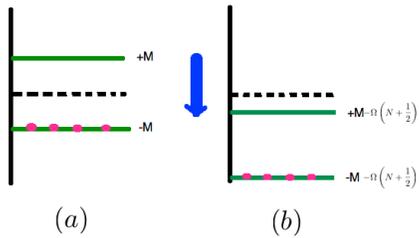}
  \caption{The particle ($+M$) and anti-particle ($-M$) LLL for $\Omega=0$ are shown in (a) each with degeneracy $N$.
  For $\Omega\neq 0$  the degeneracy is lifted. In (b) we illustrate  how the centrifugation lifts the degeneracy on the
  states with angular momentum $N$ by shifting them  down by $\pm M -\Omega(N+\frac 12)$. The rotating vacuum
  now includes the particle LLL which needs to be filled.}
    \label{fig_shift}
  \end{center}
\end{figure}

For $M=0$, (\ref{4}) exhibits a $U(2)$ symmetry as the set
$({\bf 1}, \gamma^5, -i\gamma^3, \gamma^{1+2}=-i\gamma^0\gamma^1\gamma^2)$ leaves (\ref{4}) unchanged.
This symmetry rotates particles to anti-particles. The mass upsets this symmetry, and is
 only $U(1)\times U(1)$ symmetric under the action of $({\bf 1}, \gamma^{1+2})$.   In~\cite{MIRANSKY} it was noted,
 that  for $\Omega=0$, (\ref{4}) breaks spontaneously $U(2)\rightarrow U(1)\times U(1)$ with a finite
 condensate $\left<\bar\psi\psi\right>=-N/S$ without fermionic interactions. This is readily understood from
 the illustration in Fig.~\ref{fig_shift}a, where only the LLL for particle states with spin up and mass $+M$,
 and antiparticle states with spin down and mass $-M$ are shown. Each level is $N$ degenerate. The vacuum state
 consists of filling the anti-particle states only.  Clearly, for finite $M$ the $U(2)$ symmetry is explicitly broken.
 However, as $M\rightarrow 0$ the explicit breaking is removed, but the anti-particle states remain still occupied
 eventhough they have the same zero energy as the particle states. The state breaks spontaneously the balance between
 particles and anti-particles or  $U(2)\rightarrow U(1)\times U(1)$. We now show that this {\it free} scalar  condensate
 disappears for any finite rotation $\Omega$.

For a heuristic arguments for the role of a finite rotation $\Omega$ along the magnetic field, we show
in Fig.~\ref{fig_shift}b its effect on the LLL with maximum orbital angular momentum $N$. Both the particle
and anti-particle states are shifted down and below the zero energy mark even for $M=0$. This means that
in the rotating vacuum, the particle LLL needs to be filled. Since typically the unordered scalar condensate
operator is $\bar\psi\psi\sim (a^\dagger a+b^\dagger b-1)\bar u u$, it follows for Fig.~\ref{fig_shift}b that
$\bar\psi\psi\sim (1+0-1)\bar u u=0$.

Formally, the scalar condensate carried by the rotating LLL can be explicitly constructed using the
fermionic field operator (\ref{8}). At finite temperature $1/\beta$ and $\Omega$, it is  readily found in the form

\bea
\label{16}
&&\left<\bar \psi \psi\right>(r)=\frac{eB}{2\pi}\sum \frac{e^{-\frac{eBr^2}{2}}}{m!}\left(\frac{eBr^2}{2}\right)^m \nonumber \\
&&\times (n_F(-\beta \Omega (m+1/2))+n_F(\beta \Omega (m+1/2))-1)=0\nonumber\\
\eea
which is identically zero even for zero temperature $\beta=\infty$. So any finite rotation, however infinitesimal will
cause the scalar density to vanish for free rotating fermions at finite $B$ in $1+2$ dimensions.

\subsection{Vector density}

The local density of Dirac fermions in the rotating frame in $1+2$ dimensions is readily found using
(\ref{8}) in the current density

\be
\label{10}
\left<j^0(x)\right>=\left<:\bar \psi \gamma^0\psi:\right>=\sum_{n=0}j^0_{n}(x)
\ee
The normal ordering is carried with respect to the true vacuum at finite $\Omega$.
Each LL in (\ref{10}) including the LLL
contribute through a tower of rotational states $-n<m<N-n$ for both particles and anti-particles.
This finite range in the angular momentum is further detailed in Appendix I.
Specifically, and for finite temperature $1/\beta$, the contributions of the LL and the LLL are respectively

\bea
\label{11}
j^{0}_{n>0}(x)=&&\sum_{m}|f_{nm}|^2+|f_{n-1,m}|^2 \nonumber \\
&&\times (n_F(E_{nm}^{+})-n_{F}(E_{nm}^{-}))\nonumber\\
j^{0}_{n=0}(x)=&&\frac{eB}{2\pi}\sum_{m}\frac{e^{-\frac{eBr^2}{2}}}{m!}\left(\frac{eBr^2}{2}\right)^m\nonumber \\
&&\times \frac{\sinh(\beta \Omega(m+\frac{1}{2})/2)}{\cosh (\beta \Omega(m+\frac{1}{2})/2)}
\eea
with the definition

\be
\label{12}
E^{\pm}_{nm}&&=E_n\mp \left(m-n+\frac 12\right)\Omega\nonumber\\
&&=\sqrt{eB n}\mp \left(m-n+\frac 12 \right)\Omega
\ee
We first note that the particle density is inhomogeneous in the plane and peaks at the edge of the disc
$S=\pi R^2$ under the effects of centrifugation. For small $\beta\Omega\ll 1$, i.e. small rotations or
high temperature,  the inhomogeneous particle density carried by the  LLL is

\bea
\label{15}
j^0_0|_{\Omega}(r)=&&\beta \Omega \frac{eB}{4\pi}\sum_{m}\frac{e^{-\frac{eBr^2}{2}}}{m!}
\left(\frac{eBr^2}{2}\right)^m\left(m+\frac 12\right)\nonumber \\
=&&\frac{\beta\Omega eB }{4\pi}\frac{1+eBr^2}{2}
\eea
Under the combined effect of the rotation and the magnetic field the particle density undergoes a {\it centrifuge effect}
with a maximum at the edge of the rotational plane. This effect will persist even in the presence of interactions as we will
discuss below (see Fig. ~\ref{fig_potentialT}).

The total number of particles follow from (\ref{10}-\ref{12}) by integration over $S=\pi R^2$. The results
for the LL and LLL are respectively

\bea
\label{13}
n_n=&&2\sum_{m}(n_F(E_{nm}^{+})-n_{F}(E_{nm}^{-}))\nonumber\\
n_0=&&\sum_{m}\frac{\sinh(\beta \Omega(m+\frac{1}{2})/2)}{\cosh (\beta \Omega(m+\frac{1}{2})/2)}
\eea
For small $\beta\Omega$, which is similar to small $\Omega$ or large temperature, the results in
(\ref{13}) simplify

\bea
\label{14}
n_n|_{\Omega}=&&4\beta \Omega \sum_{m}\left(m-n+\frac{1}{2}\right)\frac{e^{\beta E_n}}{(1+e^{\beta E_n})^2}\nonumber \\
=&&4\beta \Omega \left(\frac{N^2+2N}{2}-n\right)\frac{e^{\beta E_n}}{(1+e^{\beta E_n})^2}\nonumber\\
n_0|_{\Omega}=&&\frac{1}{2}\beta \Omega \sum_{m}\left(m+\frac 12\right)=\frac{\beta \Omega (N^2+2N)}{4}
\eea
We note that in $1+2$ dimensions, the LLL generates a net density at $\beta\Omega\ll 1$. For strictly zero temperature
(\ref{13}) gives the exact result

\be
\label{14X}
n_0|_{\beta=\infty}={\rm sgn(\Omega)}N
\ee
which can be understood from Fig.~\ref{fig_shift}b for $M\rightarrow 0$. Since the {\it normal ordered} density
operator $:\psi^\dagger \psi:\sim (a^\dagger a-b^\dagger b)u^\dagger u\sim (1-0)u^\dagger u$ which precisely gives $N$.
Note that for a rotation opposite to the magnetic field, the LLL shift up and above the zero energy mark. Therefore, we
have instead $:\psi^\dagger \psi:\sim (a^\dagger a-b^\dagger b)u^\dagger u\sim (0-1)u^\dagger u$ which precisely gives $-N$,
as expected from (\ref{14X}).

These observations are not restricted to only finite temperature. Indeed, at zero temperature but finite chemical potential,
the rotation induces changes in the population of the LLL. This can seen through  the substitution~\cite{FUKU,VOLO}

\be
\label{CHEM}
\beta\Omega\left(m+\frac 12\right)\rightarrow \beta\left(\mu+\Omega\left(m+\frac 12\right)\right)
\ee
in (\ref{13}), with the result

\bea
&&n_0(\mu)=N,\qquad\qquad\qquad\,\, \mu\geq -\frac{\Omega}2\nonumber\\
&&n_0(\mu)\approx  N+1+\frac{2\mu}{\Omega},\qquad -\left(N+\frac{1}{2}\right)\,\Omega\le\mu\le-\frac{\Omega}{2}\nonumber\\
&&n_0(\mu) =-N,\qquad\qquad\qquad \mu \le-\left(N+\frac{1}{2}\right)
\eea

\section{Interacting fermions in  $1+2$ }

Consider now fermions in $1+2$ dimensions interacting via 4-Fermi  interactions,
as a way to model QCD$_{1+2}$ in strong and rotating magnetic fields. The advantage of this reduction
is that it will allow for closed form results with physical lessons for QCD$_{1+3}$ dimensions,  which
even when modeled with 4-Fermi interactions is only tractable numerically. Following~\cite{CECIL,MIRANSKY}, we  now consider
$N_c$ copies of the preceding Dirac fermions, interacting via local 4-Fermi  $U(2)$ symmetric interactions

\be
\label{17}
{\cal L}_{\rm int}=\frac{G}{2}(|\bar \psi \psi|^2+|\bar \psi i\gamma^5 \psi|^2+|\bar \psi \gamma^3\psi|^2)
\ee
Standard bosonization gives

\bea
\label{18}
&&{\cal L}_{\rm int}\rightarrow -\bar \psi(\sigma+\gamma^3\tau+i\gamma^5\pi)\psi-\frac{1}{2G}(\sigma^2+\pi^2+\tau^2)\nonumber\\
\eea
with the scalar fields

\be
\label{19}
-\frac 1G\left(\sigma, \tau, \pi\right)= \left( \bar \psi \psi,\bar\psi\gamma^3\psi, i\bar\psi\gamma^5\psi\right)
\ee
For large $N_c$, (\ref{18}) can be analyzed in the leading $1/N_c$ approximation using the loop expansion for
the effective action. Explicit $U(2)$ symmetry makes the  effective ation only a function of $\sigma^2+\tau^2+\pi^2$,
so it is sufficient to search for saddle points with $\tau=\pi=0$, as others follow by symmetry.


The effective potential stemming from (\ref{18}) can be organized in three parts

\be
\label{20}
{\cal V}={\cal V}_0+{\cal V}_T=\frac {\sigma^2}{2G}+{\cal V}_{\Lambda}+{\cal V}_T
\ee
The zero temperature (vacuum) contribution from the fermion loop is

\be
\label{21}
{\cal V}_\Lambda=-\frac{N_c}{4\pi^{\frac{3}{2}}}\int_{\frac{1}{\Lambda^2}}^{\infty} \frac{ds}{s^{\frac{3}{2}}}e^{-s\sigma^2}eB \coth (eBs)
\ee
which is cut off in the UV by $1/\Lambda^2$, while the thermal contribution is

\be
\label{X24}
{\cal V}_{T}=-\frac{N_cT}{S}\sum_{j=1,-1}\sum_{n=0}^{N}\sum_{l=-n}^{N-n}\ln (1+e^{-\beta(E_n-j\Omega(l+\frac{1}{2}))})\nonumber\\
\ee
with $E_n=\sqrt{\sigma^2+2eBn}$ and $N/S=eB/2\pi$. A complementary but numerically useful approximation
to (\ref{X24}) is given in Appendix II using the proper time formalism.

\subsection{Weak coupling regime}


At zero temperature and in the absence of $B,\Omega$,  the effective potential
(\ref{20}) for the interacting Dirac fermions in $1+2$ dimensions simplifies

\be
\label{20X1}
{\cal V}\rightarrow \frac {\sigma^2}{2G}
 -\frac{N_c}{4\pi^{\frac{3}{2}}}\int_{\frac{1}{\Lambda^2}}^{\infty} \frac{ds}{s^{\frac{5}{2}}}e^{-s\sigma^2}
\ee
If we set $g=\frac{G\Lambda}{\pi}$, then (\ref{20X1}) exhibits a minimum
at $\sigma=\Lambda/g_r$ with $1/g_r=1/g-1/{g_c}$,
only for sufficiently strong coupling $g>g_c=\sqrt{\pi}$. The minimum breaks
spontaneously $U(2)\rightarrow U(1)\times U(1)$ with a finite
$\left<\bar\psi\psi\right>=-N_c\sigma/G$. The putative chargeless Goldstone mode
signals a BKT phase at any finite $N_c$.

At zero temperature and zero rotation $\Omega=0$ but with $B\neq 0$, the effective potential  (\ref{20})
 can be made more explicit by rescaling and expanding in $1/\Lambda$.
For small $\sigma$ and large $\Lambda$ the dominant contributions are

\bea
\label{X21}
{\cal V}_\Lambda= &&+\frac{N_c\Lambda^3 }{4\pi^{\frac{3}{2}}}\int_{1}^{\infty} \frac{dx}{s^{\frac{3}{2}}}
\frac{eBx}{\Lambda}\coth\left(\frac{eBx}{\Lambda}\right)\nonumber\\
&&-\frac{N_c\Lambda\sigma^2}{2\pi^{\frac{3}{2}}}+\frac{N_c\sigma^3}{3\pi}\nonumber\\
&&+\frac{N_c}{4\pi^{\frac{3}{2}}}\int \frac{ds}{s^{\frac{5}{2}}}(e^{-s\sigma^2}-1)(eBs\coth (eBs)-1)\nonumber\\
&&+{\cal O}\left(\frac{1}{\Lambda}\right)
\eea
The first contribution is independent of $\sigma$, so we will ignore it. Therefore, the vacuum contribution to the effective potential
combines the first term in (\ref{20}) and the second and third contributions in (\ref{X21})

\be
\label{X22}
\frac{{\cal V}_0}{N_c}\approx \frac{\Lambda \sigma^2}{2\pi g_r}
-\frac{eB}{2\pi}\sigma+\frac{\sigma^3}{3\pi}
\ee
In the weak coupling regime

\be
\label{X23}
0\leq \left(\frac 1{g_r}\equiv \frac 1g -\frac 1{g_c}\right)^{-1} \leq \frac {\Lambda}{eB}
\ee
 we can ignore the cubic contribution in (\ref{X22}).  A minimum of (\ref{X22}) always exists for arbitrarily
 weak coupling, with a mass gap $\sigma=\pi g_r N/S\Lambda$ and a finite chiral condensate
 $\left<\bar \psi\psi\right>=-N_cN/S(1-g/g_c)\approx -N_cN/S$. The latter is in agreement with the result for
 free Dirac fermions.  This is the phenomenon of magnetic catalysis~\cite{MIRANSKY}.

 \begin{figure}[t]
  \begin{center}
  \includegraphics[width=6cm]{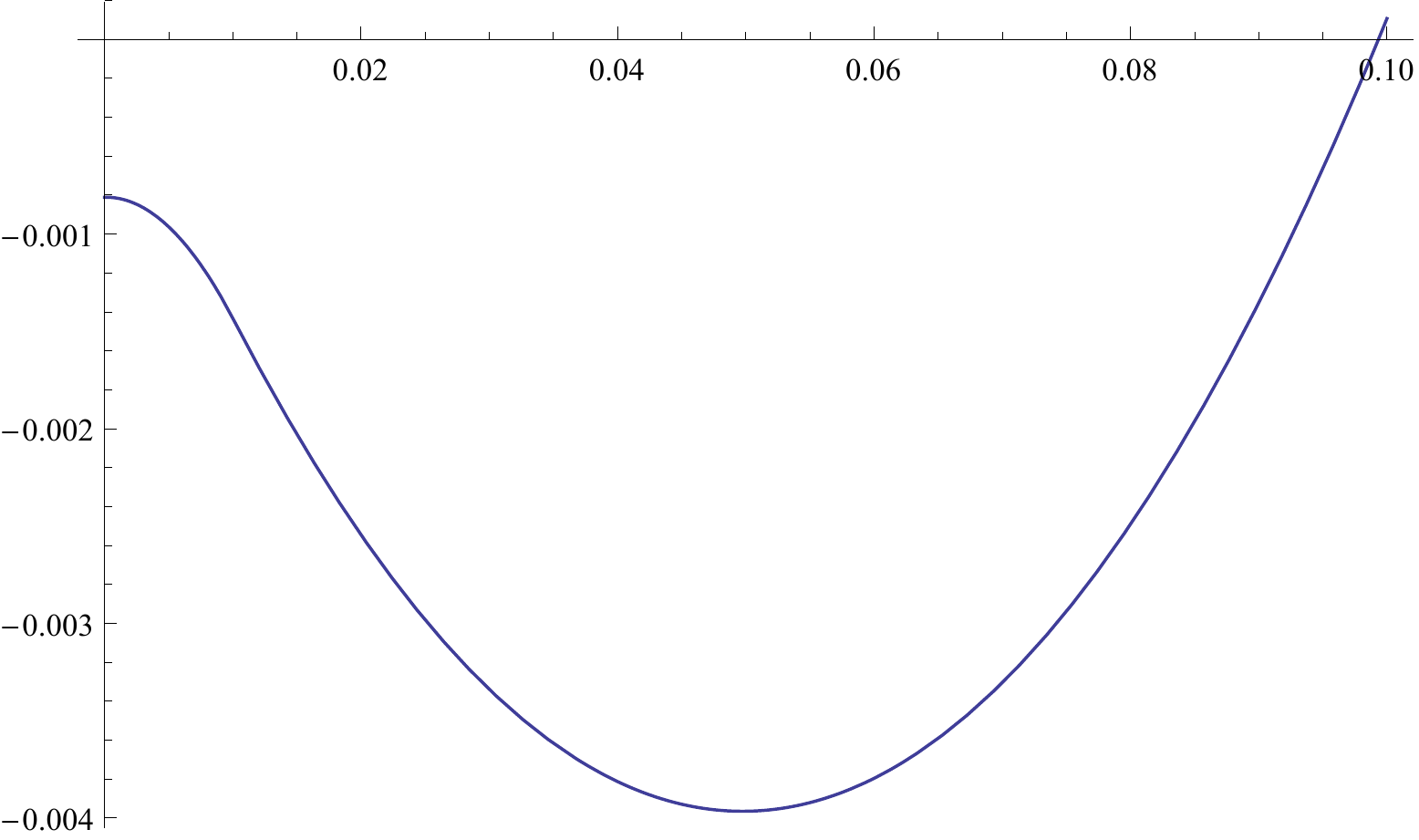}
   \includegraphics[width=6cm]{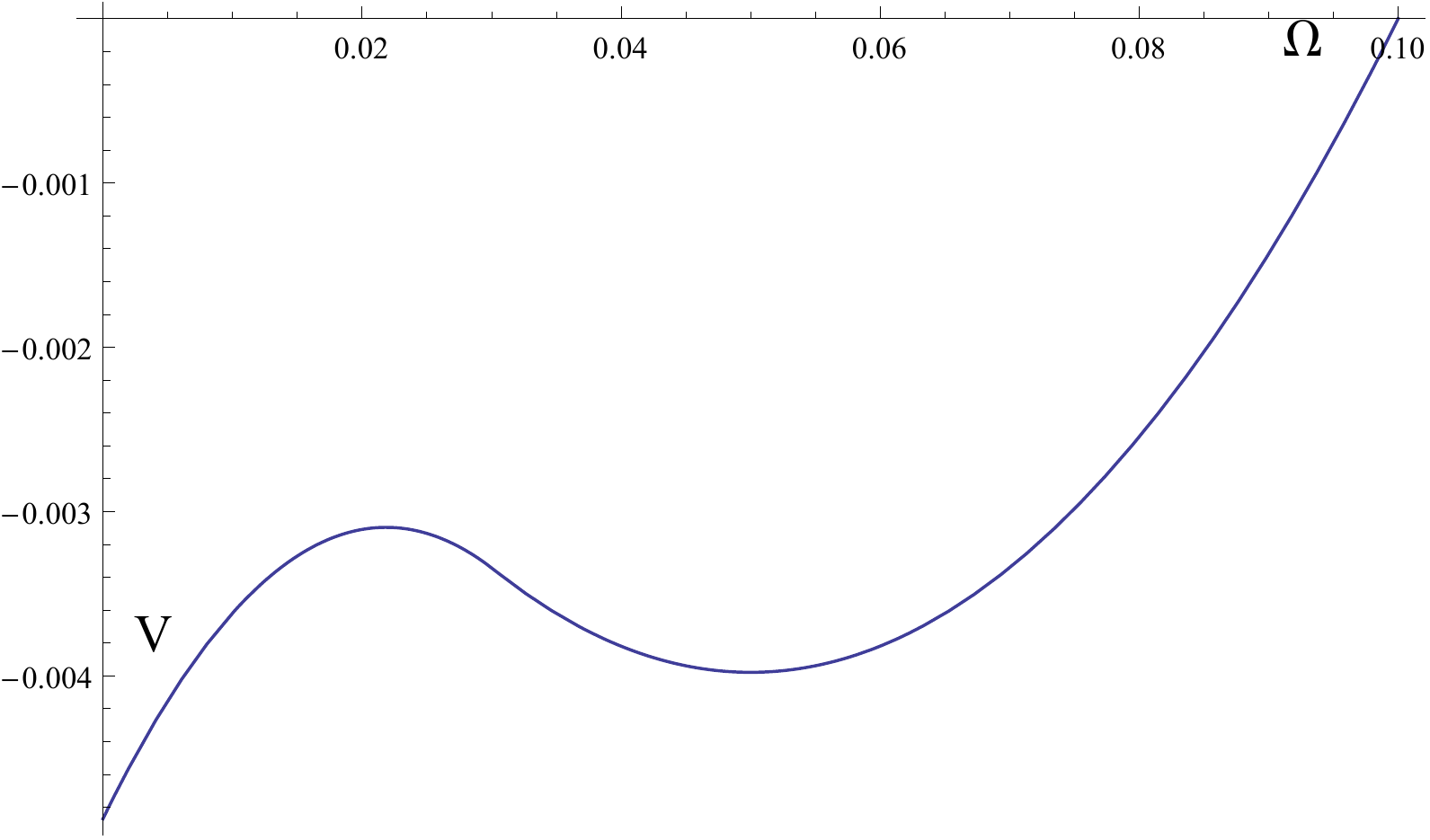}
    \includegraphics[width=6cm]{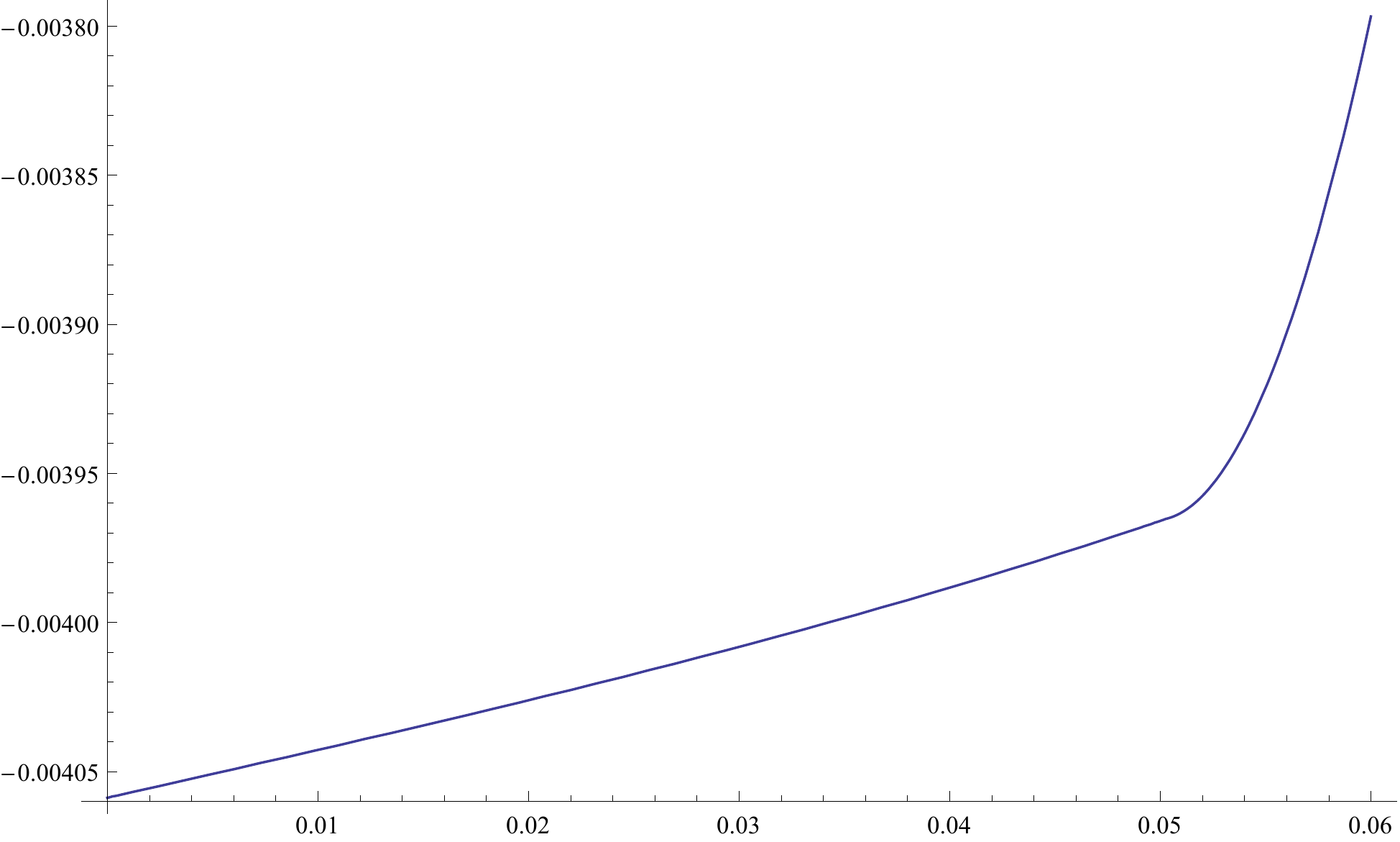}
  \caption{Effective potential ${\cal V}$ as a function of $\sigma$ in units of $\sqrt{eB}$ at $T=0$:
  $\Omega=0.0001\sqrt{eB}$ (top);  $\Omega=0.00049\sqrt{eB}$ (middle);  $\Omega=0.0005\sqrt{eB}$ (bottom).}
    \label{fig_potential0}
  \end{center}
\end{figure}

\begin{figure}[t]
  \begin{center}
  \includegraphics[width=9cm]{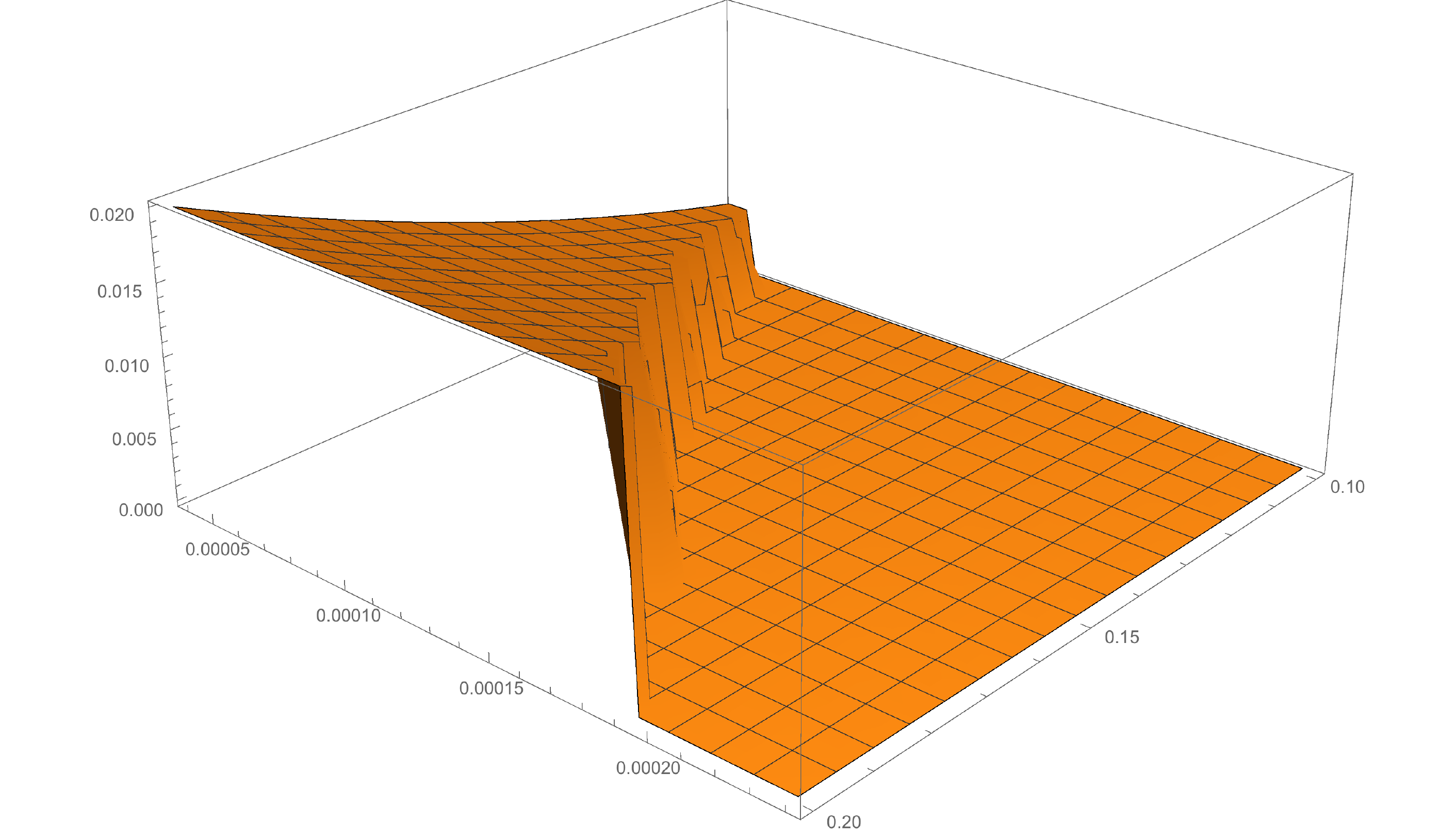}
  \caption{Effective mass as a function of $\sqrt{eB}$ and $\Omega$ in units of $\Lambda$. The
  mass gap disappears for $\Omega\geq \Omega_c$ as given by (\ref{X29X1}) through a first order transition.}
    \label{fig_potentialT=0}
  \end{center}
\end{figure}

\subsubsection{Vacuum with $\Omega\neq 0$}

At zero temperature, the effective potential for rotating Dirac particles in a strong magnetic field is given by the first two contributions
in (\ref{20}) plus the contribution from the rotating anti-particles in the LL,

\bea
\label{X25}
\frac{{\cal V}}{N_c}=&&+\frac{\Lambda\sigma^2}{2\pi g_r}-\frac{ eB}{2\pi}\sigma\nonumber \\
&&-\frac{eB}{2\pi N}\sum_{l=0}^{N}\left(\left(l+\frac 12\right)\Omega-\sigma\right)
\theta\left(\left(l+\frac 12\right)\Omega-\sigma\right)\nonumber\\
\eea
For small rotation the summation can be approximated by a continuous integration with the result

\be
\label{X26}
\frac{{\cal V}}{N_c}\approx \frac{\Lambda\sigma^2}{2\pi g_r }-\frac{ eB}{2\pi}\sigma-\frac{1}{2\Omega S}\theta(E_{\Omega}-\sigma)(E_{\Omega}-\sigma)^2
\ee
with $E_\Omega=(N+\frac 12)\Omega$. For $\sigma>E_\Omega$, the effective potential is independent
of $\Omega$, and develops a minimum for

\bea
\label{X27}
\sigma_{2}=&&+\frac{\pi g_r}{\Lambda}\frac {eB}{2\pi}\nonumber\\
\frac{{\cal V}_{2}}{N_c}=&&-\frac{\pi g_r}{2\Lambda}\left(\frac{eB}{2\pi}\right)^2
\eea
In contrast, for $\sigma<E_\Omega$, (\ref{X26})  depends  on $\Omega$ through

\be
\label{X28}
\frac{{\cal V}}{N_c}\approx \left(\frac{\Lambda }{2\pi g_r}-\frac{eB}{4\pi N\Omega}\right)\sigma^2+\frac{eB}{4\pi N}\sigma-\frac{eB\Omega}{4\pi}\left(N+\frac{1}{2}\right)\nonumber\\
\ee
and prefers always

\bea
\label{X29}
\sigma_{1}=&&0\nonumber\\
\frac{{\cal V}_{1}}{N_c}=&&-\frac {E_\Omega}2\frac {eB}{2\pi}
\eea

For $E_\Omega<\frac {\pi g_r}{\Lambda}\frac {eB}{2\pi}$ the 2-minimum (\ref{X29}) is dominant. The rotating vacuum develops a scalar condensate $\left<\bar\psi\psi\right>\neq 0$
with finite $\sigma_2$ but zero fermion density
$\left<\bar\psi\gamma^0\psi\right>=0$. In the opposite, with $E_\Omega>\frac {\pi g_r}{\Lambda}\frac {eB}{2\pi}$,
the 1-minimum (\ref{X29}) takes over. The rotating vacuum prefers a gapless solution with $\sigma_1=0$
and zero scalar condensate $\left<\bar\psi\psi\right>= 0$,
but a finite fermion density $\left<\bar\psi\gamma^0 \psi\right>\neq 0$. In large $N$, the critical value for which this occurs is

\be
\label{X29X1}
\Omega_c=\frac{g_r}{2N+1}\frac{eB}{\Lambda}
\ee
This is the phenomenon of rotational inhibition of the magnetic catalysis noted in $1+3$ dimensions in~\cite{FUKU}.
At finite but large $N$ and without the use of the continuum approximation and keeping the $\sigma^3$ term, the results remain quantitatively almost the
same, with one exception that the local minimum $\sigma_1=0$ can overtake the finite local minimum $\sigma_2$ slightlly before the $\Omega_c$. For $\Lambda=10\sqrt{eB}$ and $N=100$, (\ref{X29X1}) yields  $\Omega_c$=0.000497$\sqrt{eB}$.
We note that in the free case with $\Lambda\rightarrow\infty$, (\ref{X29X1}) yields $\Omega_c\rightarrow 0$ in agreement with
the observation in (\ref{16}).  Any finite rotation destroys the free scalar condensate.

In Fig.~\ref{fig_potential0} we show the behavior of the effective potential for finite but small $\Omega$ with the
two local minima (\ref{X27}) and (\ref{X29}). We have used $\Lambda/\sqrt{eB}=10$, $N=100$ and $g_r=1$.
A transition sets in numerically $\Omega_c=0.000488\sqrt{eB}$ in agreement with (\ref{X29X1}).
In Fig.~\ref{fig_potentialT=0} we display the  effective mass as  as function  of $ {\sqrt{eB}} $ and $\Omega$ in units of $\Lambda$,
for $g_r=1$ (weak coupling regime) and $T=0$. While the mass gap is seen to increase slightly faster than linearly with $\sqrt{eB}$
at $\Omega=0$, the effects of the rotation is to cause it to disappear at the critical value (\ref{X29X1})  through a first order transition
at weak coupling.


\subsubsection{Thermal state with $\Omega\neq 0$}

First we note that the existence of a mass gap for any finite temperature does not contradict
the Mermin-Wagner-Coleman (MWC) theorem,  since the thermal state is in a BKT phase rather
than a spontaneously broken or Goldstone phase. Having said that,
at finite temperature and weak coupling, we note that since $\sigma_2\ll \sqrt{eB}$, the temperatures of interest for the
vanishing of the mass gap, are in the low range with  $T\ll \sqrt{eB}$. Therefore, only the
$j=\pm 1$ LLL contribute in (\ref{X24}). For $T\approx T_c\approx \sigma_2$,
the potential flattens out and the centrifugation near $\sigma=0$ becomes visible leading to a small value
for the critical $\Omega_c$.

\begin{figure}[t]
  \begin{center}
  \includegraphics[width=6cm]{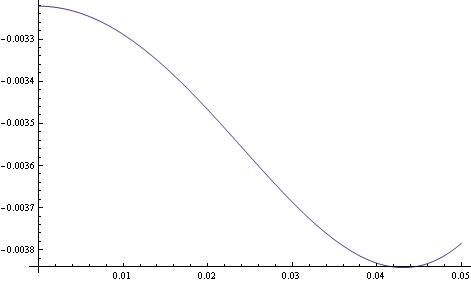}
    \includegraphics[width=6cm]{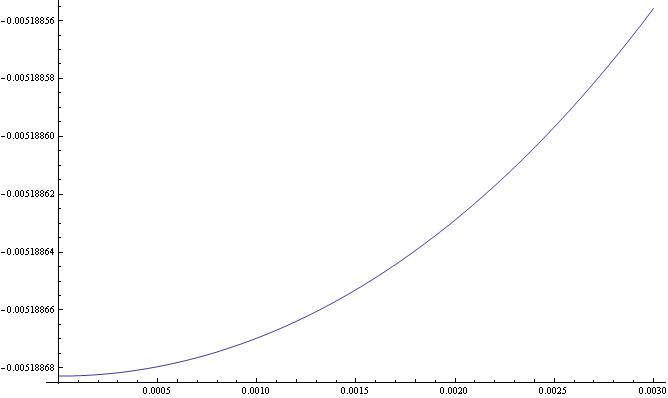}
  \caption{Finite temperature effective potential ${\cal V}$ as a function of $\sigma$ in units of $\sqrt{eB}$:
  $\beta=100/\sqrt{eB}$ and $\Omega=0.0003\sqrt{eB}$ (top); $\beta=43/\sqrt{eB}$ and $\Omega=0.0001\sqrt{eB}$
  (bottom).}
    \label{fig_potentialTT}
  \end{center}
\end{figure}

\begin{figure}[t]
  \begin{center}
   \includegraphics[width=10cm]{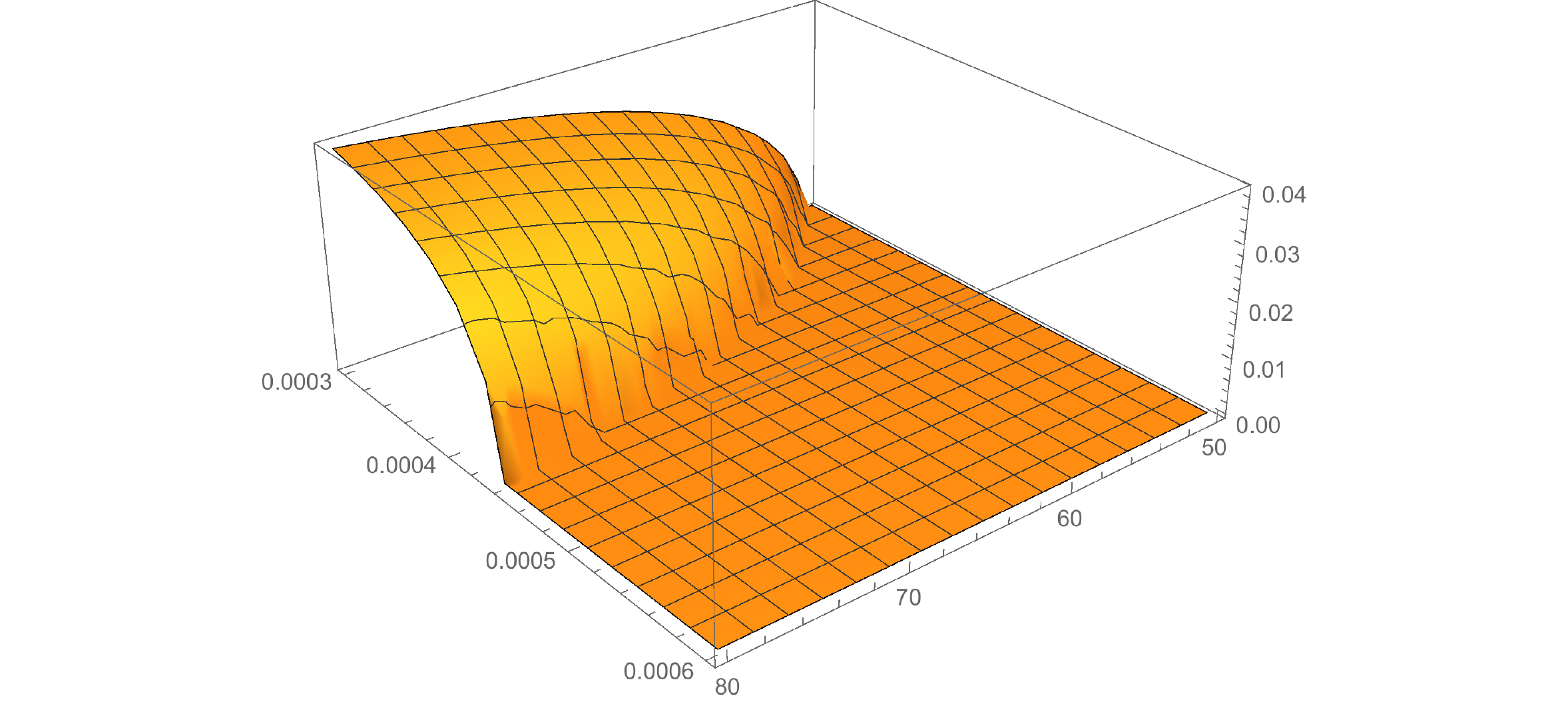}
  \caption{Effective mass as a function of $\beta$ and $\Omega$ in units of $\sqrt{eB}$ at $T\neq 0$.}
    \label{fig_potentialT3D}
  \end{center}
\end{figure}

In Fig.~\ref{fig_potentialTT} we show the behaviour of the effective potential
for  $\Lambda/\sqrt{eB}=10$, $N=100$ and $g_r=1$ (weak coupling) for $\beta=80/{\sqrt{eB}}$ and $\beta=43/{\sqrt{eB}}$.
For $\beta\geq 80/\sqrt{eB}$ the transition occurs  at $\Omega_c\approx 0.0005\sqrt{eB}$, and
for $\beta=43/\sqrt{eB}$, the transition is around $\Omega_c=0.0001\sqrt{eB}$. The critical temperature is
numerically in the range $\beta_c\approx (40-43)/\sqrt{eB}$. The behavior of the effective mass is shown in
in Fig.~\ref{fig_potentialT3D}  for the same value of $g_r=$ (weak coupling) and
${\Lambda}=10\sqrt{eB}$,  as a function of $\beta$ and $\Omega$ for the ranges $50<\beta<80$
and $0.0003\leq \Omega\leq 0.0006$ in units of $\sqrt{eB}$.

\begin{figure}[t]
  \begin{center}
  \includegraphics[width=6cm]{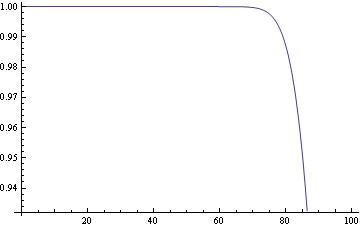}
    \includegraphics[width=6cm]{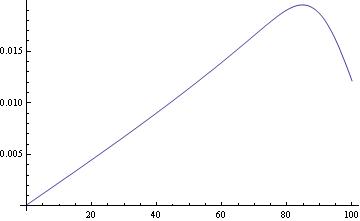}
  \includegraphics[width=6cm]{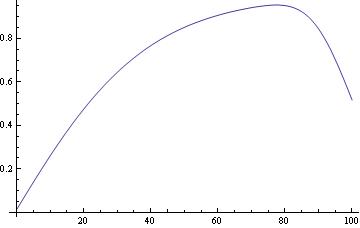}
  \includegraphics[width=6cm]{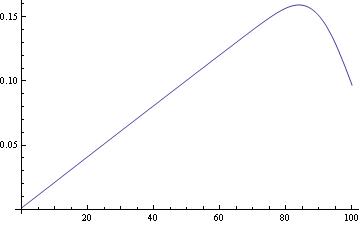}
  \caption{The current density in the weak coupling regime with $g_r=1$,
  as a function of x=$\frac{eBr^2}{2}$ in unit of $\frac{eB}{2\pi}$ at $T=0$ and $\Omega=0.0005$ (in unit of $\sqrt{eB}$) (first); $\beta=100$, $\Omega=0.0001$ (second); $\beta=100$, $\Omega=0.0005$ (third);  $\beta=40$, $\Omega=0.0001$ (fourth)}
    \label{fig_potentialT}
  \end{center}
\end{figure}

In Fig.~\ref{fig_potentialT} we show the analogue of the profile density (\ref{15})
in units of $\sqrt{eB}$, in the weak coupling regime with $g_r=1$ and for $1/\beta\ll \Omega$   as
a function of $x=eBr^2/2$. The first figure from the top is for
$\Omega=0.00005\sqrt{eB}$ for $1/\beta=0$. It is roughly constant
and drops sharply at the edge of the causality disc fixed by $\Omega R=1$.
However, for $\Omega\ll 1/\beta\ll \sqrt{eB}$ a linear behavior sets  in the middle of the
disc, to drop only sharply at the edge. The second and third figures from the top are for $\beta=100/\sqrt{eB}$  and
 $\Omega=0.0001\sqrt{eB}$ and $\Omega=0.0005\sqrt{eB}$ respectively. The fourth figure is for
$\beta=40/\sqrt{eB}$ at $\Omega=0.0001\sqrt{eB}$.  As we indicated in section IID for the free case, this
{\it centrifugation effect} holds for the interacting case as well and carries to higher dimensions as we
show below.  We will suggest a possible physical application in 1+3 dimensions. Finally, the occurence
of surface or edge modes was noted recently in~\cite{MAXIM}. We show in Appendix IX that  they do not
alter our current discussion for large $N$.

\subsubsection{Dense state with $\Omega\neq 0$}

For completeness, we now explore the effects of a finite chemical potential $\mu$ on the mass gap
for $\bar\psi \psi$ pairing.  Just as a caution, we note that a more complete treatment would require the
inclusion of the competing $\psi\psi$ channel as well. However, we note that in leading order in $1/N_c$
the $\psi\psi$ channel is $1/N_c$ suppressed in comparison to the $\bar\psi \psi$ channel and can be ignored.
With this in mind, the effect of a finite chemical potential follows from
(\ref{24}) through the substitution $\Omega(l+\frac 12)\rightarrow \mu+\Omega(l+\frac 12)$, which we now
briefly address.

\begin{figure}[t]
  \begin{center}
    \includegraphics[width=6cm]{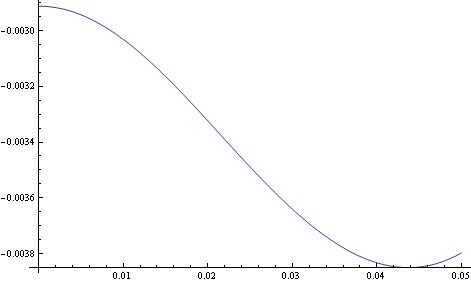}
  \includegraphics[width=6cm]{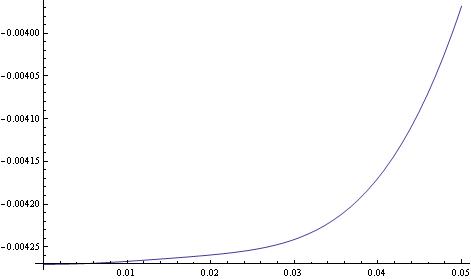}
   \caption{Finite temperature effective potential ${\cal V}(\sigma)$ at $\beta=80/{\sqrt{eB}}$
   and $\mu=0.007/{\sqrt{eB}}$ as a function of $\sigma$ in units of $\sqrt{eB}$: $\Omega=0$ (top) and
  $\Omega=0.0003\sqrt{eB}$ (bottom)}
    \label{fig_potentialTmu}
  \end{center}
\end{figure}

In Fig.~\ref{fig_potentialTmu} we show the behavior of the effective potential ${\cal V}$
for $\beta=80/\sqrt{eB}$ and $\mu=0.007/\sqrt{eB}$ as a function of $\sigma$ in units of $\sqrt{eB}$.
The top figure is for $\Omega=0$ and the bottom figure is for $\Omega=0.0003\sqrt{eB}$. The increase in the rotation
causes the loss of the gapped solution. In particular, for $g_r=1$ (weak coupling),
$\beta=80/{\sqrt{eB}}$ and $\Omega=0$,  the critical value is $\mu_c=0.02\sqrt{eB}$, while for
 $\Omega=0.0003\sqrt{eB}$, the critical value is $\mu_c=0.007\sqrt{eB}$.

 Finally and for completeness,
 we discuss in Appendix III the  dense state with {\it negative} $\mu$. Since   the model under consideration
 can be viewed as an effective description of planar condensed matter systems~\cite{CECIL}, a negative chemical
 potential is experimentally accessible.

\subsection{Strong coupling regime}

 In the opposite regime of strong coupling with $g>g_c$, a mass gap also forms.
In the regime where the ratio $\frac{\Lambda}{\sqrt{eB}}$ is large and $g>g_c$ or
$g_r<0$, the minimum of the effective potential is now controlled by the first and third contributions
in (\ref{X22}) namely

 \be
\label{X22X}
\frac{{\cal V}_0}{N_c}\approx -\frac{\Lambda \sigma^2}{2\pi |g_r|}
+\frac{\sigma^3}{3\pi}
\ee
with a mass gap $\bar\sigma=\Lambda/|g_r|$.
For $\sqrt{eB}/\Lambda<1$, the leading contribution shifts the mass and the scalar condensate quadratically,

 \be
 \label{XX24}
 \frac{\left<\bar\psi\psi\right>_B}{\left<\bar\psi\psi\right>_0}-1\approx
 \frac{(eB)^2}{12(\Lambda/g_r)^4}
 \ee
We note that the ratio of the mass gap to the LL gap $\bar\sigma/\sqrt{eB}$ can be very large.
Therefore, the critical $\Omega_c$ for which the mass gap can be depleted is much larger in strong coupling
than in weak coupling. For fixed $\Omega$ , the mass $\bar \sigma$ decreases as the ratio $\Lambda/\sqrt{eB}$ decreases.
For instance, for $g_r=-4$ and ${\Lambda}/{\sqrt{eB}}=5$, $\Omega_c \approx 0.008\Lambda$, but for  ${\Lambda}/({\sqrt{eB}|g_r}|)=3$, $\Omega_c\approx 0.009\Lambda$.
In Fig.~\ref{fig_massstrong} we show the behavior of the mass gap for strong coupling with $g_r=-4$
versus $\sigma$ in units of $\Lambda$ as a function of $\Omega$ expressed in units of $\Lambda/10^4$.
The top figure is for $\Lambda/\sqrt{eB}=5$ and the bottom figure is for $\Lambda/\sqrt{eB}=3$.

 \begin{figure}[t]
  \begin{center}
  \includegraphics[width=6cm]{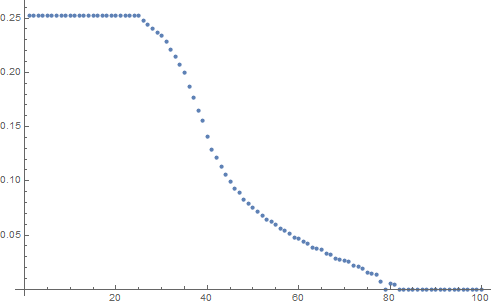}
    \includegraphics[width=6cm]{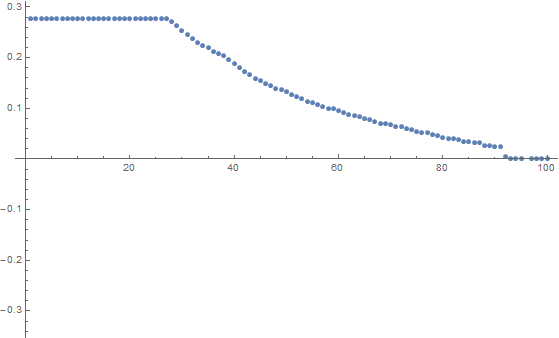}
  \caption{Mass gap $\sigma/\Lambda$ in the strong coupling regime with $g_r=-4$, as a function of $\Omega/(10^{-4}\Lambda)$
  for   ${\Lambda}/{\sqrt{eB}}=5$ (top) and ${\Lambda}/{\sqrt{eB}}=3$ (bottom).}
    \label{fig_massstrong}
  \end{center}
\end{figure}

\section{Free Dirac fermions in  1+3}

The extension of the previous analysis to 1+3 dimensions for free Dirac fermions is straightforward. In Appendix IV we detail the
rotating wavefunctions in the presence of a magnetic field, for the free case. The interacting case is more challenging for say the
case of QCD which is strongly coupled and gapped in the vacuum. Below, we will focus on the combined effects of a rotation and
magnetic field on the QCD chiral condensate in the spontaneously broken phase using mesoscopic arguments, and
leading order chiral perturbation.

\subsection{Free left currents}

We now extend the analysis for the left or L-currents  to show the generic nature of the
observations made in $1+2$ dimensions above. From Appendix IV,
the L-wavefunctions in $1+3$ dimensions  take the simplifying form

\bea
\label{23}
u_L(n=0)=&&v_l(n=0)=\sqrt{\frac{\tilde E-p}{2\tilde E}}(f_{0,m},0)\nonumber\\
u_L(n,m)=&& \frac{1}{\sqrt{2\tilde E(\tilde E+p)}}(\sqrt{2eBn}f_{nm},(\tilde E+p)f_{n-1,m})\nonumber\\
v_{L}(n,m)=&& \frac{1}{\sqrt{2\tilde E(\tilde E+p)}}(\sqrt{2eBn}f_{nm},-(\tilde E+p)f_{n-1,m})\nonumber\\
\eea
The left particle density at the origin is

\be
\label{24}
&&\frac{2\pi}{eB}n_L(0)=\nonumber\\&&+\int_{-\infty}^{0}\frac{dp}{2\pi}(n_F(-p-\mu_{00})-n_F(-p+\mu_{00}))\nonumber\\
&&+\sum _{n=1}\int_{-\infty}^{\infty}\frac{dp}{4\pi}(n_F(E_n-\mu_{00})+n_F(E_n-\mu_{10}))\nonumber \\
&&-\sum _{n=1}\int_{-\infty}^{\infty}\frac{dp}{4\pi}(n_F(E_n+\mu_{00})+n_F(E_n+\mu_{10}))\nonumber\\
\ee
while the current density at the origin is

\be
\label{25X}
j^3_L(0)=\frac{eB}{2\pi}\left(J^3_{L,0}+\sum_{n=1} J^3_{L,n}\right)
\ee
with

\bea
\label{26}
J^3_{L,0}=&&-\int_{-\infty}^{0}\frac{dp}{2\pi}(n_F(-p-\mu_{00})-n_F(-p+\mu_{00}))\nonumber \\
=&&-\frac{\Omega}{4\pi}-\frac{\mu_L}{2\pi}\nonumber\\
J^3_{L,n}=&&-\sum _{n=1}\int_{-\infty}^{\infty}\frac{dp}{4\pi}(n_F(E_n-\mu_{00})-n_F(E_n-\mu_{10}))\nonumber \\
&&+\sum _{n=1}\int_{-\infty}^{\infty}\frac{dp}{4\pi}(n_F(E_n+\mu_{00})-n_F(E_n+\mu_{10}))\nonumber\\
\eea
with $\mu_{00}=\frac{\Omega}{2}+\mu_L$ and $\mu_{10}=-\frac{\Omega}{2}+\mu_L$.
For small $B$ and zero $\mu_L$, the summation in (\ref{25X}) gives

\be
\sum \frac{eB}{2\pi}f(\sqrt{p^2+2gBn})\rightarrow \int \frac{kdk}{2\pi}f(\sqrt{p^2+k^2})
\ee
This reproduces the known result at $B=0$~\cite{VILENKIN}

\be
-\frac{T\Omega}{12\pi^2}-\frac{(\Omega+2\mu_L)^3+(\Omega-2\mu_L)^3}{96\pi^2}
\ee

While the current density at the origin reproduces the expected result, the distribution of
the current density in the radial direction is not homogeneous. Indeed, the centrifugation causes it
to peak at the edge as in $1+2$ dimensions. This is readily seen from the contribution of the LLL
which can be worked out explicitly with the result

\be
J^{3}_{Ln=0}=-\frac{eB}{4\pi^2}\sum_{m=0}e^{-\frac{eBr^2}{2}}\left(\frac{eBr^2}{2}\right)^m\frac{(m+1/2)\Omega +\mu_L}{m!}
\nonumber\\
\ee
The sum can be performed exactly with the result

\be
\label{CENT}
J^{3}_{Ln=0}(r)=\frac{eB}{4\pi^2}\left(\mu_L+\Omega\left(\frac 12 + \pi Nr^2\right)\right)
\ee
The {\it centrifugal effect} causes the current density to peak at the edge of the rotational
plane in $1+3$ dimensions.

A  possible application  of this phenomenon maybe in current heavy ion collisions
at collider energies such as RHIC and LHC. Indeed, for semi-central collisions both the rotational (orbital) and
electric magnetic fields are sizable with $\Omega\sim {eB}\sim m_\pi$ which may induce partonic densities of the type
(\ref{CENT}) that are largely deforned in the transverse plane. While the rotation and magnetic fields tend to separate
the partonic charges in concert along the rotational axis, the centrifugation causes this separation to peak in the
orthogonal direction where the observed particle flow is more important. If true, this effect should be seen as an
enhancement of $v_4$ in the charged particle flow.

\subsection{Number of free left particles}

As we noted in $1+2$ dimensions, the number of free left particles increases in
$1+3$ dimensions due to the sinking of the particle LLL in the Dirac sea. More explicitly, we have

\bea
\label{ROT0}
n_L&&=\int dx dy \left<:\bar \psi_L \gamma^{0}\psi_L:\right>\nonumber \\
&&=\sum_{m}\int_{-\infty}^{0} \frac{dp}{2\pi}(n_F(-p-\mu_m)-n_{F}(-p+\mu_{m}))\nonumber \\
&&+\sum_{n=1,m}\int_{-\infty}^{\infty} \frac{dp}{2\pi}(n_F(E_n-\mu_{nm})-n_{F}(E_n+\mu_{nm}))\nonumber\\
\eea
Here $\mu_{nm}=(m-n+\frac{1}{2})\Omega+\mu_L$ and $E_n=\sqrt{p^2+2eBn}$. The flowing left current
along the rotational-magnetic axis is

\bea
\label{ROT}
j^3_L=&&\int dx dy \left<\bar \psi_L \gamma^{3}\psi_L\right>\nonumber \\ =&&-\sum_{m}\int_{-\infty}^{0} \frac{dp}{2\pi}(n_F(-p-\mu_m)-n_{F}(-p+\mu_{m}))\nonumber \\ =&&-\frac{1}{2\pi}\sum_{m=0}^{N}\left(m+\frac{1}{2}\right)\Omega+\mu_L\nonumber \\=&&-\frac{\Omega }{2\pi}\left(N+\frac{N^2}{2}\right)-\frac{\mu_L N}{2\pi}
\eea
The first contribution in (\ref{ROT}) was noted in~\cite{YIN,FUKU}.
(\ref{ROT0}-\ref{ROT})  generalize to arbitrary $1+d$ dimensions. In particular, for  $\mu_L=0$

\be
\label{14X1}
n_{L0}=\frac{2^{\frac{d-3}2}V_{d-2}}{(2\pi)^{d-2}}{\rm sgn}(\Omega)|\Omega|^{d-2}
\sum_{m=1}^N\left(m+\frac 12\right)^{d-2}
\ee
with the volume $V_{d-2}=\pi^{\frac d2-1}/\Gamma(\frac d2)$.

\subsection{Relation to anomalies}

These observations can be used to generalize (\ref{ANO1})  to arbitrary $1+d=2n$ dimensions.
Consider the case with non-vanishing and non-parallel magnetic fields  $B_{2k,2k+1}\ne 0$ with $1\le k \le n-3$.
The general anomaly induced  chiral magnetic effect for the left current is~\cite{LOGA}

\be
\label{ANO1}
J^{2n-1}_{L\mu_L}=-\frac{\mu_L}{2\pi}\left(\frac{e}{2\pi}\right)^{n-1}B_{12}B_{34}....B_{2n-4,2n-3}
\ee
We now observe from (\ref{CENT}) that the role of the rotation is to tag to $\mu_L$  in $2n=4$ dimensions as

\be
\frac {eB}{2\pi}\left(\mu_L+\Omega\left(\frac 12 + \pi Nr^2\right)\right)\equiv
\mu_L\frac{eB}{2\pi}+\Omega\,J(r)\nonumber\\
\ee
The anomalous result (\ref{ANO1}) relates to the rotationally induced current by a similar subsitution in $2n$ dimensions,
namely

\be
\label{ANO2}
J^{2n-1}_{L\Omega}(r)=-\frac 1{2\pi}\left(\frac{e}{2\pi}\right)^{n-2}B_{12}B_{34}....B_{2n-6,2n-5}(\Omega,J(r))\nonumber\\
\ee
where $J(r)$ refers to the current spin density in the radial direction within  the $2n-4,2n-3$ plane

\be
J_{2n-4,2n-3}(r)=\frac{eB_{2n-4,2n-3}}{2\pi}\left(\frac{1}{2}+B_{2n-4,2n-3}\frac{r^2}{2}\right)
\ee
The rotational contribution to the current density (\ref{ANO2}) in 2n dimensions
 is related to the chiral magnetic effect (\ref{ANO1}) in $2n-2$ dimensions.

\subsection{Charge neutral volume}

\begin{figure}[t]
  \begin{center}
  \includegraphics[width=8cm]{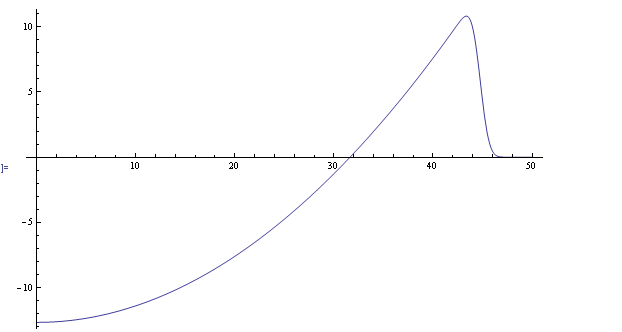}
    \caption{The charge distribution (\ref{HUMP}) in the LLL  in a closed volume $V$ with overall charge neutrality,
    for $N=1000$ as a function of $r$ and in units of $eB$.}
    \label{fig_charge}
  \end{center}
\end{figure}

Most of the analyses for the fermions presented above hold for the {\it absolute} ground state with overall charge conservation
not enforced (open volume $V$). If we require total charge neutrality of the system (closed volume $V$) then we expect an induced
charge chemical potential  $\mu_{\rm in}$ such that ($\vec\Omega\cdot \vec B>0$)

\bea
&&\sum_{n,m=0}^{N}\int \frac{dp}{2\pi}n_F\left(E_n-\mu_{\rm in}-\Omega\left(\frac{1}{2}+m-n\right)\right)=\nonumber \\&&\sum_{n,m=0}^{N}\int \frac{dp}{2\pi}n_F\left(E_n+\mu_{\rm in}+\Omega\left(\frac{1}{2}+m-n\right)\right)
\eea
where the number of $\pi^+$ (first contribution) balances the number of $\pi^-$  (second contribution).
For large $eB$ or small temperature $T$, only the $n=0$ term survives as before. In this case, the solution for 
$\mu_{\rm in}$ follows by inspection

\be
\mu_{\rm in}=-\frac{\Omega}{2}-\frac{N\Omega}{2}
\ee
The ground state consists of negative charge filling the LLL with 
$m=0$ to $m=\frac{N}{2}$, and  positive charge filling the LLL with $m=\frac{N}{2}$ to $N$. The corresponding charge density for masless  fermions is

\bea
\label{HUMP}
&&\left<J_{L,n=0}^0(x)\right>=\nonumber\\
&&\frac{eB}{4\pi^2}\sum_{m=0}^{[\frac{N}{2}]}e^{-\frac{eBr^2}{2}}\left(\frac{eBr^2}{2}\right)^m\frac{(m-\frac{N}{2})\Omega}{m!}\nonumber \\ &&+\frac{eB}{4\pi^2}\sum_{m=[\frac{N}{2}]+1}^{N}e^{-\frac{eBr^2}{2}}\left(\frac{eBr^2}{2}\right)^m\frac{(m-\frac{N}{2})\Omega}{m!}
\eea
The first line is the contribution from all negative charge contributions,  and the second line from all positive charge 
contributions. After integration, the total negative charge density  is

\be
\left<\int d^2x  J_{L,n=0}^0(x)\right>_{\rm negative}=\frac{1}{2\pi}\sum_{m=0}^{[\frac{N}{2}]}\left(m-\frac{N}{2}\right)\Omega
\ee
and similarly for the positive charge density. In Fig.~\ref{fig_charge} we  display the charge density in the
LLL in a closed volume $V=SL$ with total charge neutrality as given by (\ref{HUMP}). We expect the same
distribution of charge around a fluid vortex when overall charge neutrality is enforced, which is to be
contrasted with a vortex with only positive (negative) charge accumulation when the charge neutrality 
constrain is not enforced~\cite{YIN}.

\section{Interacting Dirac fermions in  1+3}

Now we consider the case of interacting Dirac fermions in the context of QCD in 1+3 dimensions at strong coupling.
In this regime, a mass gap forms and chiral symmetry is spontaneously broken with a triplet of charged Golstone modes.
They play the role of diffusons in the vacuum~\cite{MARIUS}. We will not quantify these statements by evaluating the QCD
vacuum energy density shift caused by a finite $\Omega, B$ and use it to extract the chiral condensate.

\subsection{Diffusion with $B,\Omega=0$}

The spontaneous breaking of the symmetry is manifest though a finite scalar condensate,
which in the chiral limit relates to the quark return probability in proper time $\tau$ as~\cite{MARIUS}

\be
\left<\bar\psi\psi\right>_{0,0}=-\lim_{m\to0}\lim_{V_4\to\infty}\frac{1}{V_4} \int_0^{\infty} P(0, \tau)\,d\tau
\label{002}
\ee
with

\be
P(0,\tau)=\left<\left|u^+(\tau) u(0)+d^+(\tau) d(0)\right|^2\right>
\label{00}
\ee
for 2 light $u,d$ flavors. The averaging in (\ref{00}) is over the QCD vacuum in Euclidean 4-dimensional space.
In the absence of magnetism, the vacuum is isospin symmetric and the correlator  in (\ref{00}) is dominated by
the lightest Goldstone modes $\pi^{0,\pm}$

\be
P(0,\tau)=2\left(P_0(0,\tau)+P_\pm (0,\tau)\right)\approx \sum_Q e^{-\tau D(0,0)Q^2 }
\label{000}
\ee
The sum is over the  pions  or diffusons  with momenta $Q_{\mu}=n_{\mu}2\pi/L$
in a periodic $V_4=L^4$ Euclidean box. The vacuum diffusion constant is
$D(0,0)=2F_\pi^2/|\left<\bar\psi\psi\right>_{0,0}|$~\cite{MARIUS}.

\subsection{Diffusion with $B,\Omega\neq 0$}

Under rotations all $\pi^{0,\pm}$ are affected by centrifugation, while only the $\pi^\pm$ are affected
by magnetism. As a result, the squared and Euclideanized pion spectra are

\bea
Q_0^2=&&p_r^2+p_3^2+\left(p_4+i\Omega l\right)^2+m_\pi^2\nonumber\\
Q_{j=\pm}^2=&&eB(2n+1)+ p_3^2+ \left(p_4+i\Omega jl\right)^2+m_\pi^2
\label{003X}
\eea
Each chargeless mode carries $l=0, \pm 1, ...$, while each charged mode is
in a LL $n$ where $-n\leq l\leq N-n$ with degeneracy $N$.
Note that the rotational energy shift in Euclidean space is purely imaginary.
The change in each of the  return probabilities in (\ref{000}) following from (\ref{003X}) is

\bea
P_0(\Omega, \tau)=&&\sum_{n_r,n_3,n_4}\sum_{l=-\infty}^{+\infty} \nonumber\\
&&\times e^{-\tau D(\Omega, B) (p_r^2+ p_3^2+ \left(p_4+i\Omega l\right)^2+m_\pi^2) }\nonumber\\
P_{j=\pm}(B, \Omega, \tau)=&&\sum_{n_3,n_4}\sum_{n=0}^N\sum_{-n\leq l\leq N-n} \nonumber\\
&&\times e^{-\tau D(\Omega, B) (eB(2n+1)+ p_3^2+ \left(p_4+i\Omega jl\right)^2+m_\pi^2) }\nonumber\\
\label{004X}
\eea
with $p_{3,4}=n_{3,4}(2\pi /L)$ in an Euclidean box of 4-volume chosen cylindrical with
$V_4\rightarrow \pi R^2L^2$ and the causal constraint  $\Omega R<1$. In general,
in the rotating vacuum with magnetism the diffusion constant $D(\Omega, B)$ is $\Omega, B$
dependent.

The change in the quark return probability is the change in the charged diffuson modes and is
captured by the difference

\be
I=\int_0^{\infty}[P(\Omega, B,\tau)-P(0,0,\tau)]d\tau
\label{005}
\ee
In the chiral limit, replacing the sums over free momenta by integrals allows to get rid of the
explicit $\Omega$ dependence in (\ref{004X}) by shiting $p_4$. So the dependence on $\Omega, B$ in
$P_0$ is only through $D(\Omega, B)$. Clearly, in the absence of $B$ a rotation $\Omega$
alone cannot change the return probability, and therefore the chiral or scalar condensate as the vacuum  is rotationally
symmetric. This is not the case in the presence of an externally fixed magnetic field $B$ as rotational
symmetry is broken. Indeed, the LL dependence in $P_\pm$ does not drop but can be resummed exactly with the result

\be
I=\frac{eBV_4}{16\pi^2D}\int_0^{\infty} \left(     \frac{1}{z \sinh z} -\frac{1}{z^2} \right)dz =
-\frac{{\rm ln}2}{16\pi^2}\frac{eBV_4}{D(\Omega, B)}\nonumber\\
\label{006}
\ee
Using the  value of the diffusion constant we arrive at

\be
\frac{\left<\bar\psi\psi\right>_{\Omega,B}}{\left<\bar\psi\psi\right>_{0,0}}-1= \frac{{\rm ln}2}{16\pi^2}\frac{eB}{F_\pi^2}
\frac{D(0,0)}{D(\Omega, B)}
\label{007}
\ee
For $\Omega=0$ and $B\neq 0$,  (\ref{007}) is in agreement  with
chiral perturbation theory in leading order~\cite{SMILGA}.  This linear magnetic catalysis is supported by lattice
simulations~\cite{LATTICE}.

(\ref{007})  is the analogue of (\ref{XX24}) in $1+2$ dimensions at strong coupling, with the difference that it
grows linearly not quadratically.  The quadratic growth follows from the absence of charged Golstone modes.
As  indicated earlier, in 1+2 dimensions the gapped phase is a BKT phase not a Goldstone phase.
We now give an  independent determination that fixes $D(\Omega, B)$ in (\ref{007}).

\subsection{Energy densities of a BEC of chiral pions}

To assess the dual action of $\vec\Omega\cdot \vec B>0$ in the QCD vacuum energy,  requires vacuum loops
in the presence of $\Omega, B$. When the magnetic field is  sufficiently weak, i.e. $|eB|\ll (4\pi F_\pi)^2$
with $F_\pi$ the pion decay constant, the loop
momenta are small and QCD is well described by an effective theory of chiral pions. In leading order, the pion
interactions which are soft can be ignored. The  $\Omega, B$ dependent parts in the QCD vacuum energy follow
from a  one-pion loop with arbitrary $\Omega, B$ insertions in leading order, with the rotation acting as an effective
chemical potential.

In the presence of a fixed magnetic field in the +z direction ${\bf B}=B\hat z$, the charged
$\pi^\pm$ pion spectrum is characterized by highly degenerate LL with energies

\be
\label{SZ1}
{E}_{np}=\left(|eB|(2n+1)+p^2+m_\pi^2\right)^{\frac 12}
\ee
Each LL $n$ for fixed pion 3-momentum $p$ carries a degeneracy $N$, labeled by
the z-component of the angular momentum $L_z=l$ with $-n\leq l\leq N-n$ as detailed in Appendix XIV.
When a rotation $\Omega$ parallel to the magnetic field is applied, the spectrum (\ref{SZ1}) shifts  so that in the
rotating frame we have ($\vec\Omega\cdot\vec B>0$)

\be
\label{SZ2}
{E}_{np}\rightarrow {E}_{np}-{\Omega}{L_z} \equiv {E}_{np}-j{\Omega}{l}
\ee
Here $j=+1$ for positively charged pions (particles) and  $j=-1$ for negatively charged pions
(anti-particles). As a result, the degeneracy of each LL is lifted.
The mechanism of $\pi^\pm$ splitting by a rotation  can cause $\pi^+$ pion
condensation~\cite{US1}. We now explore this condensation in the vacuum and also matter
for different overall charge constraints.


\subsubsection{Open volume}

We first consider the open volume $V=SL$ case, where charge is free to move in and out of $V$.
In leading order in the pion interaction, the QCD vacuum energy  per unit volume in $V$
 is the sum of a purely $B$ dependent contribution ${\cal E}_{\pi B}$ and a mixed $B,\Omega$ dependent
contribution ${\cal E}_{\pi\Omega }$

\bea
\label{008}
{\cal E}_\pi(\Omega, B)={\cal E}_{\pi B}+{\cal E}_{\pi \Omega }
\eea
If we denote by ${\bf n}$ the number of
condensed $\pi^+$ per unit length $L$ along the rotational axis, then

\bea
\label{009}
{\cal E}_{\pi B}=&&2\,\frac NS\int_{-\infty}^{+\infty}\frac{dp}{2\pi}\sum_{n=0}^\infty \frac 12 \epsilon_n(p)\nonumber\\
{\cal E}_{\pi \Omega}=&&-\frac{\bf n}{S}{(N\Omega-m_0)}+c_N\frac{{\bf n}^2}{S}
\eea
with $\epsilon_n^2(p)=p^2+m_n^2$ and $m_n^2=(2n+1)eB+m_\pi^2$.
The first contribution
stems from the pion loop with charged $\pi^\pm$ pions, while the second contribution stems from the
Bose condensation of the $\pi^+$ in the LLL when the rotationally induced chemical potential $\mu_N=\Omega N$
exceeds the effective pion mass $m_0$. In the open volume case, the accumulation of the charge
at the edge of $V$  is compensated by a deficit outside of $V$ to maintain overall charge conservation.
The last contribution in ${\cal E}_{\pi \Omega}$ is the Coulomb repulsion in the condensed droplet of $\pi^+$
by centrifugation.

To assess  the Coulomb contribution, we note that
the 2-dimensional charge distribution in this state is given by $\rho_N(\tilde x)=e^2|f_{0N}(x,y)|^2$
where $f_{0N}(x,y)$ is the N-LL

\be
f_{0N}(x,y) \approx\left( \frac 1{\sqrt{2eB}}\left(2\frac{\partial}{\partial z} +\frac{eB\overline z}2\right)\right)^N \, e^{-\frac 14 eB z\overline z}
\ee
with $z=x+iy$ and valued in $S=\pi R^2$. The condensate lies at the edge of the rotational plane with a Coulomb factor

\be
c_N=\frac {e^2}{2L} \int_{L\times S} d^3x\,d^3x^\prime\,\rho_N(\tilde x)\frac{1}{|{ x}-{x}^{\prime}|}\rho_N(\tilde x^{\prime})
\ee
In the large degeneracy $N$ limit, we can aproximate this distribution by a uniform radial distribution within the area $N-\sqrt{N}\le\frac{eB r^2}{2}\le N+\sqrt{N}$ with total charge $e$. It follows that the Coulomb factor is $c_N\approx {e^2}/{12\pi\sqrt{N}}$.

The condensate density ${\bf n}$ is fixed by  minimizing the energy density ${\cal E}_{\pi \Omega}$ in (\ref{009}),
with the result

\bea
\label{CHARGE}
{\bf n}=&&\theta(N\Omega-m_0)\frac{N\Omega-m_0}{2c_N}
\eea
for  which  the energy density in (\ref{009}) is

\be
\label{009X}
{\cal E}_{\pi \Omega}\rightarrow -\frac{3\pi\sqrt{N}}{e^2S}(N\Omega-m_0)^2\theta(N\Omega-m_0)
\ee
For $eB=0.1\,m_{\pi}^2$, and $N=1000$ , the threshold for developing non-zero ${\bf n}$ is $\Omega_{\rm min}=0.001\,m_{\pi}$. For $\Omega=0.0015\,m_{\pi}$, we have ${\bf n}=268\,m_{\pi}$, and for  $\Omega=0.002\,m_{\pi}$, we have ${\bf n}=566\,m_{\pi}$.

The condensation of charged pions by rotation in a magnetic field is for bosons,  what the accumulation
of vector charge in a vortex threaded by a magnetic field is for fermions~\cite{YIN}, and in general in any
rotating frame with a magnetic field~\cite{YIN,LIAO,FUKU,US1}. For Dirac fermions this phenomenon
is related to spectral flow  and therefore to anomalies~\cite{YIN,US1}, of which the charged pionic condensate
is its low energy manifestation in the QCD vacuum. In both cases, the charge accumulation in the finite
volume $V=LS$ is compensated by a deficiency of opposite charge in the outside of the volume $V$.
Overall charge conservation is maintained by allowing the charge to move in or out of $V$ as also suggested
in~\cite{YIN} for fermions.

\subsubsection{Closed volume}


If the volume $V=SL$ is closed  with no charge allowed to flow in or out, then charge conservation is to be
enforced strictly in $V$~\cite{US1}. Let $\mu$ be the charged chemical potential in the co-moving frame. Charge neutrality
at finite $T,\mu$ requires

\bea
\label{P1}
&&\sum_{l=0}^{N}\int\frac{dp}{2\pi}\frac{1}{e^{\frac 1T (E_{0p}-l\Omega-\mu)}-1}=\nonumber\\
&&\sum_{l=0}^{N}\int\frac{dp}{2\pi}\frac{1}{e^{\frac 1T (E_{0p}+l\Omega+\mu)}-1}
\eea
with the pion spectrum (\ref{SZ1}).
(\ref{P1})  is solved for $\mu=-\frac{N\Omega}{2}$ at any temperature $T$. Therefore,
$l=N-m$ and $l=m$  state for $\pi^{+}$ and $\pi^{-}$ will have the same ocupation number.
For $N\Omega>2m_0$  simultaneous condensation occurs for $m=0$, i.e. $\pi^+$ with  $l=N$ and $\pi^-$ with $l=0$.
For $(N-2)\Omega>2m_0$ the condensation involves both $m=0,1$. As we increase $\Omega$ all $m\leq \frac N2$
will condense, i.e. $\pi^+$ with $\frac{N}{2}\le l\le N$ and $\pi^-$ with $0 \le l\le\frac{N}{2}$.

An alternative way to see this without solving for $\mu$ is to note that for all terms in (\ref{P1}) to be meaningful,
the inequalities

\be
....\le-m_0\le\mu\le m_0-N\Omega \le...
\ee
must hold.
Thus, as long as $m_0-N\Omega<-m_0$ or $N\Omega>2m_0$ , the occupation number of the $l=N$ state for $\pi^{+}$ and
the $l=0$ state for $\pi^{-}$  are no longer meaningful, and condensation may follow.  For increasing  $\Omega $ such that $m_0-N\Omega+\Omega<-m_0-\Omega$ or $(N-2)\Omega>2m_0$, the condensation for the $l=N-1$ state of $\pi^{+}$ and
the $l=1$ state for $\pi^{-}$ will also follow, which is  consistent with the above argument based on the solution for $\mu$. We note
that in the charge-conserving case, the critical $\Omega$ is twice the critical $\Omega$ in the non-conserving case.

Now consider the rotating ground state with $T=0$ and $N\Omega>2m_0$ but $(N-2)\Omega<2m_0$, so that
only the $l=N$ state for $\pi^{+}$ and $l=0$ state for $\pi^{-}$ condense. The analogue of (\ref{009}) is then

\be
\label{P3}
{\cal E}_{\pi\Omega}=-{\bf n}\,(N\Omega-2m_0)+d_N{\bf n}^2
\ee
with the new Coulomb factor

\be
d_N\approx \frac {e^2}2  \int_{l_M}^{R} 2\pi rdr \left( \frac{1}{2\pi r}\right)^2=\frac{e^2}{4\pi}\ln \frac{R}{a}\approx \frac{e^2}{8\pi} \ln N
\ee
$d_N$ is  the electric field energy stored between two charged rings with radius $l_M\sim 1/\sqrt{eB}$ and charge $-1$ ($\pi^-$),
and radius $R\gg l_M$ and charge $+1$ ($\pi^+$). The Coulomb self-energy is now subleading as $c_N/d_N$ at large $N$ and omitted.
The pion condensate density that minimizes (\ref{P3}) is the same as (\ref{CHARGE}) with the substitution $m_0\rightarrow2m_0$
and $c_N\rightarrow d_N$.

\subsubsection{Magnetic back-reaction}

To order $\alpha=e^2/4\pi$, the charged pion condensate at the edge of the volume $V$ induces a magnetic
field that adds to the applied external magnetic field, for both the open and closed case. To assess it, consider the QED part of the charged pion Lagrangian in
leading order

\be
{\cal L}=-\frac{\bf f^2}{4}+|(d+ie(A+{\bf a}))\Pi|^2
\ee
in form notations with ${\bf f}=d{\bf a}$.
Here $A$ is the external vector potential for the background magnetic field, and {\bf a} is a  fluctuation
which is 0 in leading order. At next to leading order ${\bf a}={\bf a}[J^{\mu}]={\bf a}[{\bf n}]$, with
$J^{\mu}=\left<{\bf n}|\hat J^{\mu}|{\bf n}\right>$ the current induced by the pion condensation with

 \bea
 && |\left.{\bf n}\right>_a=(a_{p=0,n=0,l=N}^{\dagger})^{{\bf n}L}(b^{\dagger}_{p=0,n=0,l=0})^{{\bf n}L}|\left.0\right>\nonumber\\
&& |\left.{\bf n}\right>_b=(a_{p=0,n=0,l=N}^{\dagger})^{{\bf n}L}|\left.0\right>
\eea
More details regarding the quantization of free pions at finite $\Omega, B$ can be found in Appendix XIV.
The sub-label $a$ refers to the closed volume case with charge conservation, while $b$ refers to the open
volume case. For both cases, the induced current is azimuthal

\bea
\label{P10}
&&J^{\theta}[{\bf n}]=\left<{\bf n}|\hat J^{\theta}|{\bf n}\right>=\frac{e{\bf{n}}N}{m_0r}|f_{0N}|^2\nonumber \\
&&\approx \frac{eN{\bf{n}}}{2m_0\pi R^2}\delta(r-R)= \frac{e^2B{\bf n}}{4\pi m_0}\delta(r-R)
\eea
with $f_{0N}$ the LLL. (\ref{P10}) sources a uniform magnetic field in $V=SL$ in the z-direction,

\be
{\bf b}_z[{\bf n}]=\frac{e^2B{\bf n}}{4\pi m_0}
\ee
which adds to  the applied external magnetic field $B\rightarrow B+{\bf b}_z[n]$. We can solve anew the LL
problem in the modified magnetic field $B(1+\frac{e^2{\bf n}}{4\pi m_0})$, which amounts to the following
substitutions

\bea
&&m_0^2\rightarrow m_0^2[{\bf{n}}]=m_\pi^2+eB\left(1+\frac{e^2{\bf n}}{4\pi m_0}\right)\nonumber\\
&&N\rightarrow N[{\bf n}]=N\left(1+\frac{e^2{\bf n}}{4\pi m_0}\right)
\eea
In addition, (\ref{P10}) induces a magnetic energy per unit length in $V$

\be
\frac{b^2}{2}\pi R^2=\frac{{\bf{n}}^2e^4B^2R^2}{32\pi m_0^2[{\bf n}]}=\frac{e^3BN{\bf{n}}^2}{16 \pi m_0^2[{\bf n}]}
\ee

The Coulomb factors in the back-reacted case are now
 $c_N=\frac{e^2}{12\pi \sqrt{N({\bf n}})}$ (open volume) and $d_N=\frac{e^2\ln N({\bf n})}{8\pi}$ (closed volume).
 With all in mind, the pion energies per  unit volume for the closed ($a$) and open case ($b$) are respectively

\bea
\label{MBACK}
&&{\cal E}^a_{\pi\Omega}[\Omega,{\bf n}]=\nonumber \\
&&-(N({\bf  n})\Omega-2m_0({\bf  n})){\bf n}+{\bf  n}^2e^2\left(\frac{eBN}{16 \pi m_0^2[{\bf n}]}
+\frac{\ln N({\bf n})}{8\pi}\right)\nonumber\\
&&{\cal E}^b_{\pi\Omega}[\Omega,{\bf n}]=\nonumber \\
&&-(N({\bf  n})\Omega-m_0({\bf  n})){\bf n}+{\bf  n}^2e^2\left(\frac{eBN}{16 \pi m_0^2[{\bf n}]}
+\frac{1}{12\pi \sqrt{N({\bf n}})}\right)\nonumber\\
\eea
We have checked that the dependence of $m_0[\bf{n}]$ and  $N[\bf{n}]$ on $\bf{n}$
is rather weak, and the threshold for pion condensation remains the same in both cases.

\subsection{Shift in the chiral condensate}

In leading order in $(eB)/(4\pi F_\pi)^2$, the chiral condensate can be extracted from
(\ref{008}-\ref{009}) as
$\left<\bar\psi\psi\right>_{\Omega, B}= {\partial {{\cal E}_\pi}(\Omega,B)}/{\partial m}$
modulo vacuum renormalization.
Using the GOR relation $m_\pi^2 F_\pi^2=-m\left<\bar\psi\psi\right>_{0,0}$ in the absence of $\Omega, B$, we can trade the derivative with respect to the current mass $m$
 with the derivative with respect to the pion mass $m_\pi$. For the $\Omega$ independent pion contribution in (\ref{008})
we explicitly have

\be
\label{010X}
\frac{\partial {{\cal E}_{\pi B}}}{\partial m}=
\frac {\left<\bar\psi\psi\right>_{0,0}}{(4\pi F_\pi)^2}\int ds \,\frac{eB\,e^{-sm_\pi^2}}{s\,{\rm sinh}(eBs)}
\ee
The corresponding shift in the chiral condensate for $\Omega=0$ but finite $B$ is

\be
\frac{\left<\bar\psi\psi\right>_{B}}{\left<\bar\psi\psi\right>_{0,0}}-1= \frac{{\rm ln}2}{16\pi^2}\frac{eB}{F_\pi^2}
\label{007X}
\ee
in agreement with
chiral perturbation theory in leading order~\cite{SMILGA}.  This linear magnetic catalysis is supported by lattice
simulations~\cite{LATTICE}. The quadratic magnetic catalysis in NJL-type models at strong coupling, was initially proposed in~\cite{MIRANSKY}.
A rerun of the same arguments for the $\Omega$ dependent contribution in (\ref{009}), yields
the net shift of the chiral condensate for the open case (no back-reaction)

\bea
&&\frac{\left<\bar\psi\psi\right>_{\Omega,B}}{\left<\bar\psi\psi\right>_{0,0}}-1= \nonumber\\
&&\frac{1}{2}\frac{eB}{F_\pi^2}\left(\frac{{\rm ln}2}{8\pi^2}-\frac{3}{e^2\sqrt{N}}\left(\frac{N\Omega}{m_0}-1\right)\theta(N\Omega-m_0)\right)
\label{007XX}
\eea
and for the closed case (no back-reaction)

\bea
\label{007XXX}
&&\frac{\left<\bar \psi\psi\right>_{\Omega,B}}{\left<\bar \psi \psi\right>_0}-1=\nonumber \\
&&\frac{eB}{2F_{\pi}^2}\left(\frac{\ln 2}{8\pi^2}-\frac{4}{e^2N\ln N}\left(\frac{N\Omega}{m_0}-2\right)\theta(N\Omega-2m_0)\right)\nonumber\\
\eea
in leading order in the pion interaction.

Finally, the back-reacted energy densities (\ref{MBACK}) can be used to correct (\ref{007XX}-\ref{007XXX}).
A rerun of the preceding arguments yield in the closed case with back-reaction

\bea
\label{BB1}
&&\frac{\left<\bar \psi \psi\right>_{B,\Omega}}{\left<\bar \psi \psi\right>_0}-1=\nonumber\\
&&\frac{eB\ln 2 }{16\pi^2F_{\pi}^2}+\theta(N\Omega-2m_0)\frac{B}{NF_{\pi}^2m_0e}\nonumber \\
&&\times \left(\frac{8 m_0-4N\Omega}{2\ln N+\frac{eB N}{ m_0^2}}+\frac{2eBN(2m_0-N\Omega)^2}{ m_0^3(2\ln N+\frac{eB N}{m_0^2})^2}\right)
\eea
and in the open case with back-reaction

\bea
\label{BB2}
&&\frac{\left<\bar \psi \psi\right>_{B,\Omega}}{\left<\bar \psi \psi\right>_0}-1=\nonumber\\
&&\frac{eB\ln 2 }{16\pi^2F_{\pi}^2}+\theta(N\Omega-m_0)\frac{B}{ NF_{\pi}^2m_0e}\nonumber \\
&&\times \left(\frac{2m_0-2N\Omega}{\frac{4}{3\sqrt{N}}+\frac{eB N}{m_0^2}}+\frac{2eBN(m_0-N\Omega)^2}{ m_0^3(\frac{4}{3\sqrt{N}}+\frac{eB N}{m_0^2})^2}\right)
\eea

The change of the chiral condensate under the combined effects of a magnetic field and a rotation was initially suggested using
arguments from random matrix theory and anomalies~\cite{US,SIMONS}. It was clarified and detailed in the context of the NJL
model in~\cite{LIAO,FUKU}.
The effect of the rotation is to inhibit the so-called magnetic catalysis as emphasized in~\cite{FUKU}.  Note that
all the shifts  are of order
$N^{-1}_c$ and would be missed in an effective calculation with constituent quarks such as in the NJL
model in the leading loop or $N_c^0$ approximation. A
critical rotation rotation can compensate the increase induced by the magnetic field.
The shifted condensates (\ref{007XX}) (open volume), (\ref{007XXX}) (closed volume) and
(\ref{BB1}-\ref{BB2}) (back-reaction) when compared to the diffusive result (\ref{007}) fix the ratio of the
diffusion constants for the different charge conservation cases, with or without magnetic back-reaction.

\section{Pion superfluid in heavy-ion collisions}

In a heavy ion collision at collider energies, very large
angular momenta  $l\sim 10^3-10^5\,{\hbar }$~\cite{STAR} and large magnetic fields
$B\sim m_\pi^2$~\cite{DIMA} are expected in off central collisions, in the early parts
of the collision. Assuming that they persist in the freeze-out part where the constituents
are hadrons, i.e. $R\sim 10$ fm with still $eB\sim m_\pi^2$, this would translates to a LL degeneracy
$N=eBR^2/2\sim (m_\pi\times10\,{\rm fm})^2\sim 100/4$ and a rotational chemical potential $\mu_N=N\Omega\sim 1.25\,m_\pi$.
The  pion chemical potentials at freeze-out are $\mu_f\sim 0.5\,m_\pi$ at RHIC,  and $\mu_f\sim 0.70\,m_\pi$ at the LHC~\cite{BEGUN}. When combined with the rotationally induced chemical potential, we have
 $\mu_\pi=\mu_N+\mu_f\sim 1.75\,m_\pi$  and $1.96\,m_\pi$  respectively. These chemical potentials may
 induce charged pion condensation, in the form of a rotating BEC of pions ad the edge of the fire ball.
 The specifics of this BEC depends on whether the volume $V$ is open or closed as we now detail.


In the open volume case without magnetic back-reaction, the mean number of condensed $\pi^+$ is

\be
\mathbb N_+=\frac{\sum_{n=0}^{\infty}n\,e^{n\frac{N\Omega+\mu_f-m_0}{T}-\frac{n^2}{12\pi\sqrt{N}TL}}}
{\sum_{n=0}^{\infty}e^{n\frac{N\Omega+\mu_f-m_0}{T}-\frac{n^2}{12\pi\sqrt{N}TL}}}
\ee
For  $L\sim 10$ fm, $eB\sim m_\pi^2$ and  $N\approx 25$, we show in Fig.~\ref{fig_massstrong} the average number of
condensed $\pi^+$  for temperatures in the range $0.5\,m_\pi\le T\le 1.5\,m_\pi$ and rotations in the range
$0.02\,m_{\pi}\le \Omega\le 0.0\,5\,m_{\pi}$. 
As $\Omega$ exceeds the critical $\Omega_{\rm min}$,
the number of $\pi^+$ increases.

\begin{figure}[h]
  \begin{center}
\includegraphics[width=9cm]{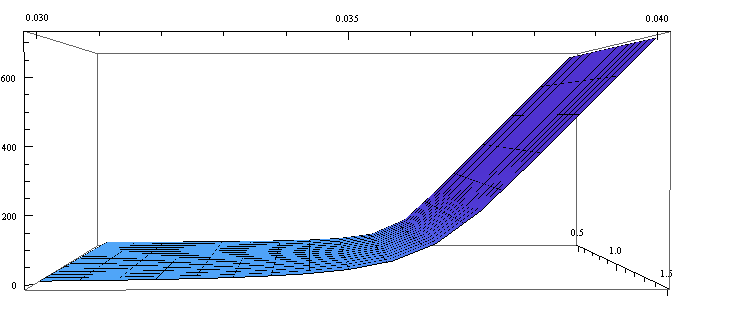}
\caption{The mean number of superfluid pions ${\mathbb N}_{\pi^\pm}$ in the range  $0.5\,m_\pi\le T\le 1.5\,m_\pi$,  $\mu_f=0.5\, m_{\pi}$ and $0.03\,m_{\pi}\le \Omega\le 0.0\,4\,m_{\pi}$.}
    \label{fig_massstrong}
 \end{center}
\end{figure}

\begin{figure}[h]
  \begin{center}
\includegraphics[width=9cm]{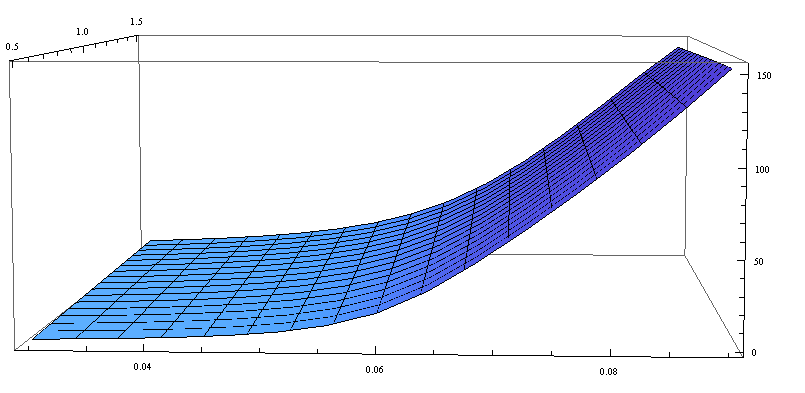}
\includegraphics[width=9cm]{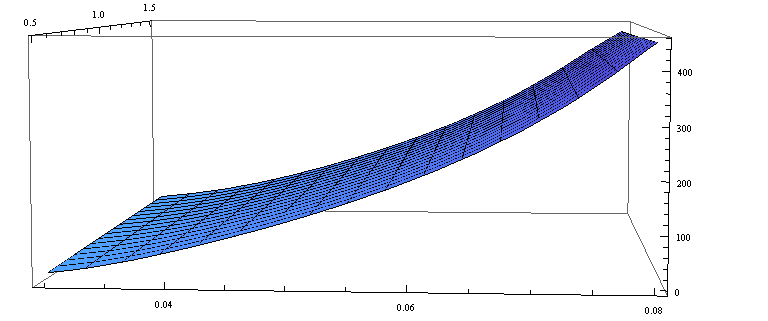}
\caption{The mean number of superfluid pions ${\mathbb N}_{\pi^\pm}$ in the range  $0.5\,m_\pi\le T\le 1.5\,m_\pi$ and
 $0.03\,m_{\pi}\le \Omega\le 0.0\,5\,m_{\pi}$, for case a (upper) and case b (lower)}
    \label{fig_massxtrong}
 \end{center}
\end{figure}

For the closed volume case without magnetic back-reaction,
the mean number of condensed $\pi^\pm$ pions  are

\be
{\mathbb N}_\pm=\frac{\sum_{n=0}^{\infty}n\, e^{\frac{n(N\Omega+2\mu_f-2m_0)}{T}
-\frac{n^2\ln N}{8\pi TL}}}{\sum_{n=0}^{\infty} e^{\frac{n(N\Omega+2\mu_f-2m_0)}{T}-\frac{n^2\ln N}{8\pi TL}}}
\ee
For  $eB=m_{\pi}^2$,  $\Omega_c=\frac{2\sqrt{2}}{N}\sqrt{eB}$ and $R\sqrt{eB}=\sqrt{2N}$, so that $\Omega_cR=\frac{4}{\sqrt{N}}$. In this case, we must have $N\ge 16$ for the critical rotation to be within the causality bound. In Fig.~\ref{fig_massxtrong} we show
${\mathbb N}_\pm$ for $N=50$  and  $L=10\, {\rm fm}\approx 7\, m_{\pi}^{-1}$, in the range $0.5\, m_{\pi}\le T\le 1.5 \,m_{\pi}$ and $0.04\, m_{\pi}\le\Omega\le 0.08\, m_{\pi}$.

\begin{figure}[h]
  \begin{center}
\includegraphics[width=9cm]{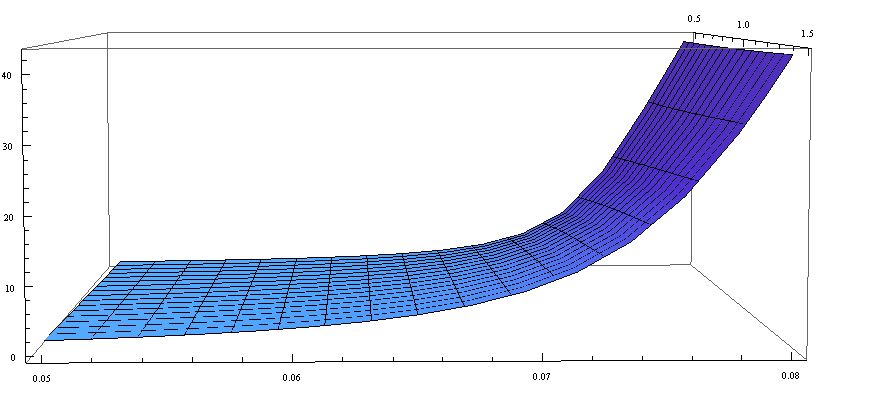}
  \caption{The mean number of superfluid pions ${\mathbb N}_{\pi^+}$ in the range  $0.5\,m_\pi\le T\le 1.5\,m_\pi$ and
 $0.03\,m_{\pi}\le \Omega\le 0.0\,8m_{\pi}$, for $\mu_f=0.8\,m_{\pi}$.}
    \label{fig_massstrong}
 \end{center}
\end{figure}


When the magnetic back-reaction is taken into account for both the closed
($a$) and open ($b$) volume case, the mean number of condensed pions is

\be
\mathbb N_{+,a,b}=\frac{\sum_{n=0}^{\infty}n \,e^{-\frac 1T(L{\cal E}^{a,b}_{\pi\Omega}[\Omega,\frac{n}{L}]-\kappa_{a,b}n\mu_f)}}{\sum_{n=0}^{\infty}e^{-\frac 1T(L{\cal E}^{a,b}_{\pi\Omega}[\Omega,\frac{n}{L}]-\kappa_{a,b}n\mu_f)}}
\ee
with $\kappa_a=2$ (closed volume) and $\kappa_b=1$ (open volume).
Below we plot the number of the condensation for $N=50$, $L=10\,{\rm fm}$, in the range of $0.03\,m_{\pi}\le\Omega\le 0.09\,m_{\pi}$.

\section{Conclusions}

We  analyzed the combined effects of a rotation and a magnetic field on free and interacting Dirac
fermions in 1+2 dimensions. Our results show that the rotation causes massless positive states in the LLL
to sink into the Dirac sea, followed by an increase in the density of particles. The scalar density of particles
does not change in the free case, but is modified in the interacting case. These results  strenghten our
earlier observation that an increase in the density of composite fermions in the quantum Hall effect at
half filling under rotation would signal their Dirac nature~\cite{US}. They may also be of relevance to planar condense
matter systems when  subject to a parallel rotation plus a magnetic field.

We showed that the mechanism behind the sinking of the LLL for free Dirac fermions,
holds in any dimension, leading to a finite increase in the density of particles that is
related to anomalies. For QCD in the spontaneously broken phase with Dirac fermions,
this mechanism manifests itself in a novel way through the condensation of charged pions. We
used this observation to derive the shift in the chiral condensate in leading order in the pion
interaction.

On a more speculative way in QCD, the charged separation caused by the dual combination of a rotation
parallel to a magnetic field, may impact on the flow of charged particles in semi-central collisions
of heavy ions at present collider energies, provided that the magnetic field is still strong in the freeze-out region.
While both the rotation and the magnetic field separate
charges along the rotational axis as known through the standard chiral vortical and magnetic effect,
the combined effect causes them to centrifuge. The resulting charge separation is quadrupolar as
opposed to polar with some consequences for the charged particle flow.   Also, the possibility of
an induced and coherent charge accumulation by rotation in a magnetic field, whether in the form of
partons or pions,  may affect  the fluctuations in the charge and pion number,
the transport coefficients such as the viscous coefficients, and potentially
the electromagnetic emissivities in the prompt and intermediate
part of the collision, especially their distribution and flow in the low mass region. These
issues are worth further investigations.

\section{\label{acknowledgements} Acknowledgements}

We thank Edward Shuryak for a discussion, and Maxim Chernodub for bringing his work to
our attention. This work was supported in part by the U.S. Department of Energy under Contract No.
DE-FG-88ER40388.


\section{Appendix I: range of $l$}

To better understand the nature of the range in the orbital angular momentum $l$ for each LL, we  recall that for
$l\geq 0$ the wavefunction is typically of the form

 \be
 \label{A1}
 z^{l}e^{-\frac{eBr^2}{4}}L_{n}^{l}(eBr^2/2)
 \ee
 The requirement that (\ref{A1}) stays within the area $S=\pi R^2$ implies that $l+n<N$, meaning that both $l,n<N$.
 Conversely, for $l<0$ the wavefunctions are of the form

\be
\label{A2}
z^{|l|}e^{-\frac{eBr^2}{4}}L_{n-|l|}^{|l|}(eBr^2/2)
\ee
which requires $n\leq N$. But for this case, we always have $n\geq -l$. These observations imply that the
orbital angular momentum is bracketed with $-n\leq l\leq N-n$. This range of $l$ helps keep the
angular  shift $\Omega n$ smaller than  the magnetic shift $\sqrt{eBn}$ for large $n$. Indeed, this requirement
together with the causality bound $\Omega R<1$, implies that

\be
\label{A3}
\sqrt{2eBn}-\Omega |l|\geq \frac{1}{R}(\sqrt{4N^2}-N)\approx\sqrt{NeB}
\ee


\section{Appendix II: alternative  ${\cal V}_T$}

The one-loop finite temperature contribution to the effective potential relates to the
scalar condensate through

\be
\label{22}
\frac{\partial {\cal V}_T}{\partial \sigma}=-\int d^2x \left<\bar \psi \psi\right>|_{\beta}.
\ee
Using the quantized fields (\ref{8}) and the proper time construction, we have

\bea
\label{23}
&&\frac{\partial {\cal V}_T}{\partial \sigma}=-4\sigma\int \frac{d\omega}{2\pi}\sum_{l}f_{F}(\omega,l) \nonumber \\
&&\times{\rm Im}\int_{0}^{\infty} ids e^{-is(\omega^2-\sigma^2-i\epsilon)}\left(\sum_{n_{min}}(2-\delta_{n,0})e^{i2eB ns}\right)\nonumber\\
\eea
For positive $l,$ the constraint is $l\le N-n$, thus the uper bound for l is $N$ and for a given l the upper bound for n is $N-l$. For negative $l$, we also have$|l|\le N$ and $ |l|\le n\le N$. Thus, the summation over $n$ gives for positive $l$

\be
\label{24}
\frac{1+e^{2ieBs}}{1-e^{2ieBs}}-2\frac{e^{2ieBs(N-l)}}{1-e^{2ieBs}}
\ee
Since we have
\be
\label{25}
f_{F}(\omega,l)=\frac{\theta(\omega)}{e^{\beta(\omega-\Omega(l+1/2)-\mu})}+\frac{-\theta(\omega)}{e^{\beta(-\omega+\Omega(l+1/2)+\mu})}
\ee
it is clear that $|f_{F}|\le 2$. Thus the summation of  the  second term in (\ref{24}) is of order

\be
\label{26}
\frac{1-e^{2ieB(Ns)}}{1-e^{2i eBs}}
\ee
After analytical continuation to the imaginary axis, this contribution vanishes in  the thermodynamical limit. The only contribution is
to the residue which is $l$-independent.
For negative  $l$ we have

\be
\label{27}
\frac{e^{2iNeBs}-e^{2iNeBs|l|}}{1-e^{2ieBs}}
\ee
After analytic continuation, neither the residue nor the integrand part survive. With all in mind, the result is now

\bea
\label{28}
\frac{\partial {\cal V}_T}{\partial \sigma}=&&-4\sigma\int \frac{d\omega}{2\pi}\sum_{l=0}^{N}f_{F}(\omega,l) \nonumber \\
&&\times {\rm Im}\int_{0}^{\infty} ids e^{-is(\omega^2-\sigma^2-i\epsilon)}\nonumber\\&&\times \left(\frac{1+e^{2ieBs}}{1-e^{2i eBs}}-2\frac{e^{2ieBs(N-l)}}{1-e^{2ieBs}}\right)
\eea
For $\omega^2-\sigma^2\le 0$, the analytical continuation of the integrand to the positive imaginary axis yields zero imaginary part.
For  $\omega^2-\sigma^2 \ge 0$  the analytical continuation of the first and second contributions to the negative and positive
real axis respectively, yield adding residues with a net imaginary part. The result is

\bea
\label{29}
\frac{\partial {\cal V}_T}{\partial \sigma}=&&-4\sigma\int \frac{d\omega}{eB}\theta(\omega^2-\sigma^2)\sum_{l=0}^{N}f_{F}(\omega,l)\nonumber \\  &&\times \left(\frac{1}{2}+\sum_{n=1}^{\infty}\cos \left(\frac{\pi n}{eB}(\sigma^2-\omega^2)\right)\right)
\eea
which integrates to

\bea
\label{30}
&&{\cal V}_T=\int d\omega \sum_{l=0}^{N}f_{F}(l,\omega)\theta(\omega^2-\sigma^2) \nonumber \\
&&\times\left(\left(\frac{\omega^2-\sigma^2}{eB}\right)+\frac{2}{\pi}\sum_{n=1}^{\infty}\frac{\sin (\frac{\pi n}{eB}(\omega^2-\sigma^2))}{n}\right)
\eea
Through a change of variable, we can recast each $l$-contribution in (\ref{30}) in the form

\bea
\label{31}
&&\int f_{F}(l,\omega)\theta(\omega^2-\sigma^2)\frac{\omega^2-\sigma^2}{eB}\nonumber \\
&&-\frac{2}{\pi}\sum_{n=0}^{\infty}\int_{\omega^2-\sigma^2\ge 2eBn} ^{\omega^2-\sigma^2\le 2eB(n+1)}f_{F}(\omega)\frac{\frac{\pi(\omega^2-\sigma^2)}{eB}-(2n+1)\pi}{2}\nonumber \\
\eea
By partial integration we found that the first  term  cancels the last term, with only boundary terms left. The final result for
the thermal contribution to the effective potential takes the canonical form

\be
\label{32}
{\cal V}_T=\frac{1}{\beta}\sum_{l=N}^{\infty}\sum_{n=0}^{\infty}\sum_{j=1,-1}\ln (1+e^{-\beta(E_n-j(\mu+\Omega(l+\frac{1}{2}))})
\ee
This result is equivalent to  (\ref{X24}) in the thermodynamical limit.

\begin{figure}[t]
  \begin{center}
  \includegraphics[width=6cm]{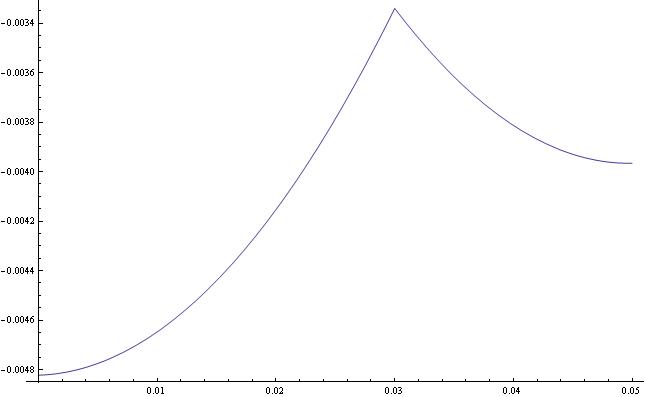}
   \includegraphics[width=6cm]{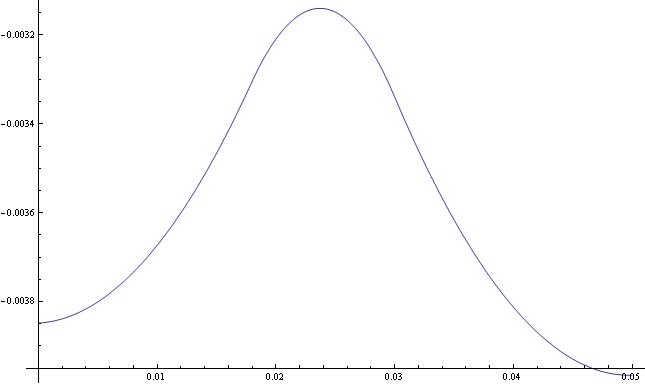}
    \includegraphics[width=6cm]{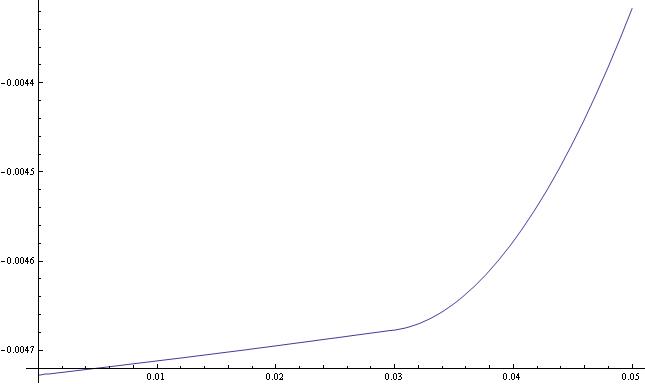}
  \caption{Effective potential ${\cal V}(\sigma)$ at $T=0$ and
  $\mu=-0.03 {1}/{\sqrt{eB}}$  in units of $\sqrt{eB}$:
  $\Omega=0$ (top); $\Omega=0.00012\sqrt{eB}$ (middle); and $\Omega=0.001\sqrt{eB}$  (bottom).}
    \label{fig_potentialTN1}
  \end{center}
\end{figure}

\begin{figure}[t]
  \begin{center}
  \includegraphics[width=6cm]{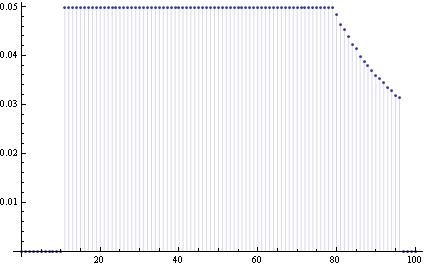}
  \caption{Effective mass at $T=0$ and
  $\mu=-0.03 {\sqrt{eB}}$ in units of ${\sqrt{eB}}$ as a function of $\Omega$ in units of $10^{-5}\sqrt{eB}$}
    \label{fig_potentialTN2}
  \end{center}
\end{figure}

\section{Edge modes  in 1+2}

Recently, it was  noted in~\cite{MAXIM}  that for a negative fermion mass and when the boundary condition at the luminal radius $R$
was enforced (for example through an MIT bag boundary condition, see also~\cite{JAP}), there is one imaginary solution to the radial wave number $k_{\perp}=\sqrt{E^2-M^2}$ for each angular momemtum $m$  (in the infinite area case $k_{\perp}^2=2eBn$). These solutions were
referred to as  edge modes as they peak near the edge in the absence of a magnetic field.  For a finite magnetic field,
 the corresponding wave function reads

 \be
 e^{-\frac{eB r^2}{4}}\,r^{m}e^{i m\phi}\,{}_1F_1\left(-\frac{k_{\perp}^2}{2eB},m+1,\frac{eBr^2}{2}\right)
 \ee
 The increasing hypergeometric function ${}_1F_1$ may overcome the pre-factor $e^{-\frac{eBr^2}{4}}\,r^{m}$,  and become dominant  at large $r$. However, for large degeneracies with $N\gg 1$ this does not take place. Indeed, in the parameter range discussed here with $N=100$ and $M=-\sqrt{eB}$, the edge solution for $m=0$ reads $k_{\perp}^2\approx -10^{-42}eB$ and for  $k_{\perp}$ this small, the hypergeometric function remains almost constant  for all $r$. This  edge mode is simply the deeply confined LLL mode $e^{-\frac{eB r^2}{4}}$. For $m=80$, the edge solution is about $k_{\perp}^2\approx -0.4\,eB$. The
${}_1F_1$ function for this value at the edge is about 2 times  the value at the origin or $r=0$, which should be viewed as a moderate enhancement of the LLL wave function with  ${}_1F_1$    set to 1. Specifically, $r^m e^{-\frac{eB r^2}{4}}$ for $m\approx N$ already peaks near the boundary, the egde ehancement by ${}_1F_1$ changes nothing qualitatively. For large $N$, the   LLL wave function remains a good aproximation for the low lying modes and needs no further amendment. The only effect is that the energy of these edge states become slightly lighter (for the case considered  it is 0.8:1), which could turn to a moderate statistical enhancement.

\section{Negative $\mu$ in 1+2}

The use of a negative potential $\mu$ maybe more than academic in 1+2 dimensions,
since effective descriptions of planar condense matter systems are
described by the model we presented in the main text using Dirac fermions~\cite{CECIL}. In Fig.~\ref{fig_potentialTN1}
we show the behavior of the effective potential ${\cal V}$ as a function of $\sigma$ for $T=0$ and $\Omega=0$, but
large negative $\mu=-0.031\sqrt{eB}$, where the gap solution is lost (top). The critical value for which this happens is
$\mu_c=-0.025\sqrt{eB}$. Amusingly, with increasing $\Omega$, the mass gap is recovered at $\Omega_{c1}$, then lost at $\Omega_{c_2}$. For instance, at  $T=0$ and $\mu=-0.03\sqrt{eB}$, we have  $\Omega_{c1}=0.00011\sqrt{eB}$ and
$\Omega_{c_2}=0.00096\sqrt{eB}$ as illustrated in Fig.~\ref{fig_potentialTN1} middle and bottom respectively.
In Fig.~\ref{fig_potentialTN2} we show the effective mass as a function of $\Omega$  for also $T=0$ and $\mu=-0.03\sqrt{eB}$.

\section{Free Dirac fermion in 1+3}

In $1+3$ dimensions, the rotating metric  (\ref{1}) is minimally changed to $ds^2\rightarrow ds^2-dz^2$, with the
pertinent changes to the co-moving coordinates. In the chiral Dirac basis for the gamma matrices, the rotating
LL levels (\ref{6}) are now changed to

\be
\label{18}
(E^{\pm}+\Omega(m-n+\frac{1}{2}))=\pm\sqrt{p^2+M^2+2eBn}=\pm \tilde E\nonumber\\
\ee
with the corresponding wavefunctions for particles

\bea
\label{19}
u^T_{nm1}=&&e^{-iE^+t+ipz}\frac{1}{\sqrt{2\tilde E(\tilde E+p)}}\nonumber \\
&&\times(M f_{nm},0,(\tilde E+p)f_{nm},-\sqrt{2eBn}f_{n-1,m})\nonumber\\
u^T_{nm2}=&&e^{-iE^+t+ipz}\frac{1}{\sqrt{2\tilde E(\tilde E+p)}}\nonumber \\
&&\times(\sqrt{2eBn}f_{nm},(\tilde E+p)f_{n-1m},0,Mf_{n-1,m})\nonumber\\
\eea
and anti-particles

\bea
\label{20}
v^T_{nm1}=&&e^{-iE^-t-ipz}\frac{1}{\sqrt{2\tilde E(\tilde E+p)}}\nonumber \\
&&\times(M f_{nm},0,-(\tilde E+p)f_{nm},-\sqrt{2eBn}f_{n-1,m})\nonumber\\
v^T_{nm2}=&&e^{-iE^-t-ipz}\frac{1}{\sqrt{2\tilde E(\tilde E+p)}}\nonumber \\
&&\times(\sqrt{2eBn}f_{nm},-(\tilde E+p)f_{n-1m},0,Mf_{n-1,m} )\nonumber\\
\eea
The  quantized fields are now

\be
\label{21}
\psi(t,\vec x)=\int \sum_{mni}\frac{dp}{2\pi }\left(e^{-iE^{+}t+ipz}u_{nmi}(x_{\perp})a_{nmi}(p)\right.\nonumber \\
\left.+e^{-iE^{-}t-ipz}v_{nmi}(x_{\perp})b^{\dagger}_{nmi}(p)\right)
\ee
with the anti-commutation rules

\be
\label{22}
[a_{nmi}(p),a_{pqj}(p^{\prime}]_+=\delta_{np}\delta_{mq}\delta_{ij}2\pi\delta(p-p^{\prime})
\ee

\section{Free pion in $1+3$}

We now present and explicit derivation of the pion spectrum in a rotating frame for infinire volume.
The rotating metric is the same as for the Dirac fermions
 in $1+3$ dimensions. The co-moving frame is defined similarly with  $e_a=e^{\mu}_{a}\partial_{\mu}$ and
$(e_0,{\bf e})=(\partial_t+y\Omega\partial_x-x\Omega\partial_y,{\bf \nabla})$. In the rest frame, the circular
 vector potential reads $A_R=-\frac {Br^2_R}2\,d\theta_R$ in form notation. Using the coordinate transform
 to the rotating frame $r_{M}=r,t_M=t,\theta_{M}=\theta+\Omega t$ yields

\bea
A=-\frac{B r^2}{2}d\theta-\frac{\Omega Br^2}{2}dt
\label{A4}
\eea
In the rotating frame there is in addition to the magnetic field $B\hat z$, an induced electric field
$\vec E=\Omega B\vec r$. This is expected from a Lorentz transformation from the fixed frame with
$B\hat z$ to the co-moving frame $B\hat z$ and $\vec E=\Omega B\vec r$.

In the rotating frame, a charged scalar is described by the Lagrangian

\be
\label{A5}
{\cal L}=&&|(D_t+y\Omega D_x-x\Omega D_y)\Pi|^2-|D_i\Pi|^2-m_\pi^2\Pi^{\dagger}\Pi\nonumber\\
\ee
with the long derivative $D=\partial +ieA$.  The electric field drops out in (\ref{A4}), thanks to the
identity

\be
\label{A55}
D_t+y\Omega D_x-x\Omega D_y=\partial_t+y\Omega \partial_x-x\Omega \partial_y
\ee
The co-moving
frame corresponds only to a frame change with no new force expected.
In the rotating frame, the charged field  satisfies

\be
\label{A6}
-(\partial_t+y\Omega \partial_x-x\Omega \partial_y)^2\Pi-D_i^{\dagger}D_i\Pi+m_\pi^2\Pi=0
\ee
In the infinite volume case, we solve (\ref{A6}) using the ladder operators

\bea
\label{A7}
a=&&\frac{i}{\sqrt{2eB}}(D_x+iD_y)\nonumber\\
a^{\dagger}=&&\frac{i}{\sqrt{2eB}}(D_x-iD_y)\nonumber\\
b=&&\frac{1}{\sqrt{2eB}}(2\partial +\frac{eB}{2}\bar z)\nonumber\\
b^{\dagger}=&&\frac{1}{\sqrt{2eB}}(-2\bar\partial +\frac{eB}{2}z)\nonumber\\
\eea
Hence,  the identities

\bea
\label{A8}
&&D_x^{\dagger}D_x+D_y^{\dagger}D_y=eB(2a^{\dagger}a+1)\nonumber\\
&&L_z=i(-x\partial_y+y\partial_x)=b^{\dagger}b-a^{\dagger}a
\eea
The general stationary solution to (\ref{A6}) is of the form $\Pi=e^{ipz-iEt}\,f$ with $f$ solving

\be
\label{A9}
(E+\Omega L_{z})^2\,f=(m_\pi^2+p^2)\, f+eB(2a^{\dagger}a+1)\, f
\ee
The normalizable solutions form a tower of LL of the form

\bea
\label{A10}
&&f_{mn}=\frac{1}{\sqrt{m!n!}}(a^{\dagger})^n(b^{\dagger})^{m}f_{00}\nonumber\\
&&(E_{mn}+\Omega(m-n))^2=eB(2n+1)+m_\pi^2
\eea
with $f_{00}\sim  e^{-\frac{eB}4(x^2+y^2)}$ as the LLL.
Therefore, the quantized charged field $\Pi$ in the rotating frame takes the form

\be
\label{A11}
\Pi= \int \frac{dp}{2\pi }\sum_{nm}\frac{f_{mn}}{\sqrt{2\tilde E_n}}(a_{nmp}e^{-iE^{+}t+ipz}+b^{\dagger}_{nmp}e^{iE^{-}t-ipz})\nonumber\\
\eea
with the bosonic canonical rules

\be
\label{A12}
  \left[b_{nmp},b^{\dagger}_{n^{\prime}m^{\prime}p^{\prime}}\right]=\left[a_{nmp},a^{\dagger}_{n^{\prime}m^{\prime}p^{\prime}}\right]=2\pi\delta_{nn^{\prime}}\delta_{mm^{\prime}}\delta(p-p^{\prime})\nonumber\\
\ee
$a^{\dagger}_{nmp}$ creates a $\pi^+$ with energy $E^{+}=E_n-\Omega(m-n)$ , charge $+e$ and
$l=m-n$. $b^{\dagger}_{nmp}$ creates a $\pi^-$ with
 energy $E^{+}=E_n+\Omega(m-n)$ , charge $-e$ and  $l=-m+n$.  Hence,
the relation between the rotating frame and the rest frame energies are
$E^{rotating}=E^{rest}-\Omega L_z$ with $L_z=jl$, $l=m-n$. In particular,
 $j=+1$ for $\pi^+$ (particle) and  $j=-1$ for $\pi^-$ (anti-particle) as in (\ref{SZ2}).
For completeness, the solutions to the Klein-Gordon equation can be found in~\cite{US1}.


 \vfil

\end{document}